\documentclass[prd,aps, preprintnumbers, showpacs,twocolumn, nofootinbib,superscriptaddress,notitlepage]{revtex4-1}
\usepackage[normalem]{ulem}
\usepackage{epsfig}
\usepackage{amsfonts}
\usepackage{amsmath}
\usepackage{slashed}
\usepackage{graphicx}
\usepackage{color}
\usepackage{mathtools}
\usepackage{subfigure}
\usepackage{graphicx}
\usepackage{float}

\usepackage{stmaryrd}
\usepackage{amssymb,psfrag}
\usepackage[normalem]{ulem}
\usepackage{caption}
\newcommand{\beq}{\begin{eqnarray}}
\newcommand{\eeq}{\end{eqnarray}}

\newcommand{\nn}{\nonumber}

\def\red{\textcolor[rgb]{1.00,0.00,0.00}}

\begin{document}

\title{An analysis of nuclear parton distribution function based on relative entropy}

\author{Shu-Man Hu}
\email{hushm2025@lzu.edu.cn}
\affiliation{Frontiers Science Center for Rare Isotopes, and School of Nuclear Science and Technology, Lanzhou University, Lanzhou 730000, China}

\author{Ao-Sheng Xiong}
\email{Corresponding author. xiongash21@lzu.edu.cn}
\affiliation{Frontiers Science Center for Rare Isotopes, and School of Nuclear Science and Technology, Lanzhou University, Lanzhou 730000, China}

\author{Ji Xu}
\email{Corresponding author. xuji@lzu.edu.cn}
\affiliation{Frontiers Science Center for Rare Isotopes, and School of Nuclear Science and Technology, Lanzhou University, Lanzhou 730000, China}

\author{Fu-Sheng Yu}
\email{Corresponding author. yufsh@lzu.edu.cn}
\affiliation{Frontiers Science Center for Rare Isotopes, and School of Nuclear Science and Technology, Lanzhou University, Lanzhou 730000, China}

\author{Ji-Xin Yu}
\email{Corresponding author. yujx18@lzu.edu.cn}
\affiliation{Frontiers Science Center for Rare Isotopes, and School of Nuclear Science and Technology, Lanzhou University, Lanzhou 730000, China}

%\date{\today}
\begin{abstract}
  In this work, we propose a method to quantify the difference between nuclear parton distribution functions in different nuclei and parton distribution functions in free nucleons using the relative entropy (also known as Kullback-Leibler divergence), a measure widely employed in quantum information theory. By introducing certain constraints and the “minimum relative entropy"  hypothesis, we can determine the shape of the structure function in the intermediate-$x$ region, which is intimately connected with the renowned EMC effect. For quark structure functions, our results align with the latest global fits to experimental data. This agreement suggests that the relative entropy-based methodology may provide novel insight into the structure of nucleons, particularly in cases where experimental data and theoretical QCD constraints are limited, such as those pertinent to gluon nPDFs. Therefore, we applied this methodology to gluon nPDFs, analyzing the results from two commonly used global fitting groups, EPPS21 and nNNPDF3.0. Our analysis suggests that the central values of EPPS21  align more closely with the “minimum relative entropy” hypothesis. This finding underscores the utility of the proposed method and provides a valuable reference for future global fitting of nPDFs.

\end{abstract}

\maketitle

%%%%%%%%%%%%%%%%%%%%%	
\section{Introduction}
\label{Introduction}
%%%%%%%%%%%%%%%%%%%%%

% \red{In this article, we propose the minimum Kullback–Leibler (KL) divergence (or relative entropy) hypothesis. By applying the minimum KL divergence criterion to the structure function related to quark nPDFs in the intermediate-$x$  region with minimal constraints, we successfully recover the established modification pattern. Building on this validation, we then apply the same method to the less biased gluon distribution. The resulting gluon nPDF shows remarkable consistency with current EPPS21 global fits, demonstrating the predictive power of the information-theoretic approach. Our findings suggest that the KL divergence, as an implementation of a minimal-information-update hypothesis, can serve as a novel and insightful observable for studying nucleon parton structure. Present the comment in the abstract.}

Parton distribution functions (PDFs) are a kind of quantum correlation functions which characterize the probability that the parton carries a certain momentum fraction of the hadron momentum. The PDFs serve as fundamental inputs for precision Standard Model (SM) baseline predictions at the LHC and other collider facilities, thus their determination is a key focus in high-energy physics \cite{Hou:2019efy,Bailey:2020ooq,Lai:2010vv,Martin:2009iq,Alekhin:2012ig,Barone:2000tx}.

When free nucleons become bound within atomic nuclei, nuclear parton distribution functions (nPDFs) become necessary. One of the primary goals of nuclear physics is achieving a comprehensive description of the structure of atomic nuclei. The nPDFs, which describe the motion of partons in the nuclear medium, remain one of the least understood aspects of nuclear structure. A prime example is the unexpected EMC effect, which indicates the quark PDFs in bound nucleons are modified compared with free ones in the intermediate-$x$ region \cite{EuropeanMuon:1983wih,NewMuonNMC:1990xyw,Gomez:1993ri,Seely:2009gt}. The complex nature of nonperturbative QCD renders the analytical solution of nPDFs unattainable. Furthermore, lattice QCD also remains impractical for this issue, as nuclear structure represents a quintessential many-body problem that currently exceeds the computational capabilities and resources available for lattice simulations. Even for very light nuclei, current lattice calculations remain at an exploratory stage \cite{Detmold:2020snb,Amarasinghe:2021lqa,Chen:2024rgi}. Therefore, nearly all current available nPDFs are extracted from global fitting analysis of experimental data \cite{Eskola:2021nhw,AbdulKhalek:2022fyi,Duwentaster:2022kpv,Helenius:2021tof}. 

Moreover, our knowledge of gluon nPDFs is considerably more limited. While modern global fitting analyses yield increasingly precise quark nPDFs---with nuclear modifications for quarks in the intermediate-to-large $x$ region being broadly consistent across different fitting groups---the gluon nPDF remains substantially more uncertain. In contrast to the quark sector, predictions for gluons from different groups often differ markedly. Therefore, developing new methodologies to pin down parton (especially gluon) nPDFs is a significant scientific goal.

The recent rapid development of quantum information theory has drawn attention in high-energy physics for its potential applications to long-standing problems, possibly revealing fundamental connections to underlying physical mechanism \cite{Galindo:2001ei,ELEMENTSOFINFORMATIONTHEORY}. Quantum information theory has introduced novel observables---various types of entropies that enable an extended characterization of hadron structure or scattering processes beyond classical approaches. Studies in this area have proliferated significantly \cite{Wang:2014lua,Kharzeev:2017qzs,Beane:2018oxh,Gotsman:2020bjc,Fedida:2022izl,Hentschinski:2022rsa,Liu:2022grf,Han:2020vjp,Ma:2018wtw,Hagiwara:2017uaz,Vedral:2002zz,Nam:2025hei,Miller:2023ujx,Zhang:2023oja,Bloss:2025ywh,Hu:2025jne,Wang:2025inr,Lin:2025eci}. Here, we want to find an appropriate entropy for the investigation of nPDFs. First, both PDFs and nPDFs are probability distributions. Second, we want to quantify the difference between these two distributions. The appropriate quantity is the Kullback-Leibler (KL) divergence $\int p(x) \ln (p(x)/q(x))dx$ (also known as relative entropy) \cite{Kullback:1951zyt,Wehrl:1978zz,Benito-Calvino:2022kqa,Cao:2022iqh,Cao:2022ajt}, which is a type of statistical distance: a measure of how much a true distribution $p(x)$ is different from a reference  distribution $q(x)$. If we treat the PDF as the reference distribution and the nPDF as the true distribution, their ratio reflects the nuclear modification, whose fundamental mechanism may be recast in the quantum-theoretic terms of KL divergence. When a free nucleon is placed inside an atomic nucleus, its PDFs are modified into nPDFs due to interactions with surrounding nucleons. Understanding the precise mechanism of this transformation remains a challenge. This paper explores an information-theoretic approach to this problem, elaborating on how concepts such as the “minimum relative entropy" hypothesis can shed new light on this transformation.

In this paper, we quantify the difference between PDFs and nPDFs by KL divergence. In the quark sector, we determine the nuclear structure function in the intermediate-$x$ (EMC) region, where it is primarily sensitive to the underlying quark nuclear modifications, by invoking the minimum relative entropy hypothesis. This hypothesis proposes that when a free nucleon is bound in a nucleus, its PDFs transform into nPDFs in a way that minimizes the relative entropy between them, subject to known physical constraints. This approach frames the transformation as an optimization problem with fixed boundary conditions, providing a principle to determine  shape of nPDFs. In turn, this methodology offers a potential reference for assessing the quality of global fits to gluon nPDFs,  which will be elaborated below.

The organization of this paper is as follows. In Sec.\,\ref{structurefunctions}, we give the formalism for structure functions in deeply inelastic scattering (DIS). Sec.\,\ref{KLdivergence} presents the definition of KL divergence and discusses its properties. In Sec.\,\ref{numericalAnaly}, we develop a framework and present results for quantifying and determining quark structure function using the KL divergence. We turn to the investigation of gluon nPDFs in Sec.\,\ref{gluonnPDFs}. We finish with the conclusion  in Sec.\,\ref{concludingremarks}.

%%%%%%%%%%%%%%%%%%%
\section{Formalism for structure functions in DIS}
\label{structurefunctions}
%%%%%%%%%%%%%%%%%%%
The cross sections for neutral-current ($eN\to eX$) DIS on unpolarized nucleons can be written in terms of the structure functions
\begin{eqnarray}\label{startingpoint}
  \frac{d^2 \sigma_\textrm{DIS}}{d x  d y} \!=\! \frac{4 \pi \alpha^2}{Q^4} s\Big( (1-y)F_2(x,Q^2) + y^2 x F_1(x,Q^2) \Big)  \,,
\end{eqnarray}
where $x$ is the Bjorken scaling variable, $y$ denotes the inelasticity, $Q^2$ represents the negative square of four-momentum transfer from lepton to hadron, and $s$ stands for the square of the center-of-mass energy. It should be noted that Eq.\,(\ref{startingpoint}) only considers the neutral-current photon exchange at leading order (LO) in the strong coupling constant $\alpha_s$, and this equation can be further reduced by Callan-Gross relation $F_2=2x F_1$ \cite{Callan:1969uq}
\begin{eqnarray}\label{startingpoint2}
  \frac{d^2 \sigma_\textrm{DIS}}{d x  d y} = \frac{2 \pi \alpha^2}{Q^4} s\Big( 1+(1-y)^2 \Big) F_2(x,Q^2)  \,.
\end{eqnarray}
This simplified form in Eq.\,(\ref{startingpoint2}) is adopted in this study to facilitate the subsequent analysis using the KL divergence. At LO, the connection between the structure function $F_2$ and the quark distributions is given by
\begin{eqnarray}
  F_2(x,Q^2) = \sum_q e_q^2 \, x \Big[ f_q(x,Q^2) + f_{\bar q}(x,Q^2) \Big] \,,
\end{eqnarray}
where $e_q$ denotes the quark electric charge and $f_q (\bar{f}_q)$ represents the PDF of (anti)parton $q$. When considering the case of atomic nucleus, the structure function becomes
\begin{eqnarray}\label{SFforA}
  F_2^A(x,Q^2)/A &=& \sum_q e_q^2 \, x \Big[ \frac{Z}{A} \left( f_q^{p/A}(x,Q^2) + f_{\bar q}^{p/A}(x,Q^2) \right) \nn\\
  && + \frac{N}{A} \left( f_q^{n/A}(x,Q^2) + f_{\bar q}^{n/A}(x,Q^2) \right) \Big] \,,
\end{eqnarray}
where $Z$ is the atomic number, $N$ is the neutron number, and $A$ is the mass number of the nucleus. $f_q^{p/A}$ represents the nPDFs of parton $q$ in a proton
bound in nucleus A, and $f_q^{n/A}$ corresponds to a neutron.

We adopt the following notations for concision,
\begin{eqnarray}
  &&\left\{
  \begin{aligned}
  &f_q^A(x,Q^2) = \frac{Z}{A} f_q^{p/A}(x,Q^2) + \frac{N}{A} f_q^{n/A}(x,Q^2) \,,\\\\
  &f_{\bar q}^A(x,Q^2) = \frac{Z}{A} f_{\bar q}^{p/A}(x,Q^2) + \frac{N}{A} f_{\bar q}^{n/A}(x,Q^2) \,.
  \end{aligned}
  \right.
\end{eqnarray}
Therefore, the structure function for nucleus A can be simplified
\begin{eqnarray}\label{SFforA2}
  F_2^A(x,Q^2)/A &=& \sum_q e_q^2 \, x \Big[ f_q^A(x,Q^2) + f_{\bar q}^A(x,Q^2) \Big] \,.
\end{eqnarray}
Numerous experimental and theoretical studies have been dedicated to describing this structure function \cite{ParticleDataGroup:2024cfk,collinsbook,Hen:2016kwk,Radyushkin:2022qvt,Zhang:2018diq,Frankfurt:2011cs,Chen:2020ody,Chen:2015tca,Wei:2016far}. Although we do know lots of information about it, we still cannot obtain the exact distributions from first principles. The currently available and best results come from global fits of experimental data. For instance, the EPPS21 Collaboration presents the numerical values of $f_q^A$ and $f_{\bar q}^A$ \cite{datasetofEPPS21}, which show good agreement with experimental results \cite{NewMuon:1995cua,NewMuon:1995tgs,NewMuon:1996yuf,NewMuon:1996gam,EuropeanMuon:1992pyr}.

The EMC effect describes a deviation from unity in the ratio of per-nucleon deep inelastic structure functions for nucleus A to deuteron $(F_2^A(x,Q^2)/A)/(F_2^d(x,Q^2)/2)$ in the intermediate-$x$ region, which in turn may leave a signature on quantum information-theoretic quantities such as the KL divergence, and serve as a proxy for the nuclear modifications to the nucleon structure.

%%%%%%%%%%%%%%%%%%%
\section{The Kullback-Leibler divergence}
\label{KLdivergence}
%%%%%%%%%%%%%%%%%%%
In this work, we deploy the KL divergence between PDFs and nPDFs to assess the nuclear modifications. For the sake of being self-contained, we start with a short introduction of the KL divergence. Mathematically, the KL divergence between two probability distributions is defined as
\begin{eqnarray}\label{defiKL}
  D_{\mathrm{KL}}(p \| q) = \int p(x) \ln\frac{p(x)}{q(x)} dx \,.
\end{eqnarray}
From here on, we will denote $q(x)$ as the reference distribution and $p(x)$ as the true distribution. Here we employ the KL divergence, which is also referred to as relative entropy in literature, with units of \textit{nats} (with the base of the logarithm $e$).

KL divergence is a fundamental quantity in probability and quantum information theory. It arises as an expected logarithm of the likelihood ratio of two distributions $p(x)$ and $q(x)$. In what follows, we outline its distinguishing features that the KL divergence is a measure of the ``distance'' between the reference probability function $q(x)$ and the true probability function $p(x)$. It is always a non-negative real number, with value 0 if and only if the true function is identical to the  reference  function. The nonnegativity of KL divergence implicitly assumes a prerequisite condition: both functions $q(x)$ and $p(x)$ are normalized to unity, $\int q(x) dx=1,\, \int p(x) dx=1$.
%\begin{itemize}
  % \item The KL divergence is a measure of the ``distance'' between the reference probability function $q(x)$ and the true probability function $p(x)$. It is always a non-negative real number, with value 0 if and only if the reference function is identical to the true function. The nonnegativity of KL divergence implicitly assumes a prerequisite condition: both functions $q(x)$ and $p(x)$ are normalized to unity, $\int q(x) dx=1,\, \int p(x) dx=1$.

  % \item The KL divergence is not symmetric and the value from reference distribution $q(x)$ to true distribution $p(x)$ differs from that of $p(x)$ to $q(x)$, i.e., $D_{\mathrm{KL}}(p \| q) \neq D_{\mathrm{KL}}(q \| p)$.

  % \item The KL divergence can be viewed as generalization of mutual information $I(x,y)$, which is the relative entropy between a joint distribution $p(x,y)$ and the product distribution $p(x)p(y)$. The KL divergence can also be regarded as generalization of Shannon information entropy as taking the reference distribution $q(x)$ to the uniform one.
% \end{itemize}

As discussed above, the KL divergence is a useful tool for quantifying the difference between nuclear and free-nucleon distributions. In the quark sector, our analysis is formulated at the level of structure functions which are linear combinations of quark and antiquark nPDFs. In the gluon sector, the same idea is applied directly to the gluon nPDFs. We now turn to these two applications in the following sections.

%%%%%%%%%%%%%%%%%%%
\section{Framework and results for quark structure functions}
\label{numericalAnaly}
%%%%%%%%%%%%%%%%%%%
In this section, we provide calculation method of the KL divergence for different nuclei. The EMC experiment measured the ratio of per-nucleon structure functions between a nucleus A and the deuteron $(F_2^A(x,Q^2)/A)/(F_2^d(x,Q^2)/2)$, which is closely related to the nuclear modification of quark nPDFs. Therefore, determining the shape of this ratio in the EMC region provides important constraints on the corresponding quark distributions. Theoretically, a major advantage of employing this ratio lies in its insensitivity to higher-order perturbative QCD corrections, as the perturbative contributions are largely irrelevant to the nonperturbative PDFs and nPDFs. Here, we take the numerator of this ratio as the true distribution $p^A(x)$ and the denominator as the reference distribution $q^d(x)$. In this work, $Q^2=10\,\text{GeV}^2$, which is a typical scale in DIS processes. To keep the notation concise, the explicit $Q^2$ dependence is implicitly omitted.

For the reference distribution, deuteron is regarded as an approximation to a free proton and neutron system since it is loosely bound and the average distance between the nucleons is large. Therefore, we adopt the CT18A free PDFs to construct the deuteron reference structure function \cite{Hou:2019efy}. For the true distribution, we take values from the EPPS21 parametrization for different kinds of nuclei \footnote{https://research.hip.fi/qcdtheory/nuclear-pdfs/epps21/} \cite{Eskola:2021nhw}. First, we evaluate the KL divergence over a  full-$x$ interval. Because this interval includes the shadowing, antishadowing, EMC, and Fermi-motion regions, it provides a useful overall comparison of the nuclear and deuteron structure functions, and serves as supplementary background for the subsequent analysis in the EMC region. Since we perform numerical integration of global fitting results, we cannot directly compute the KL divergence over the idealized full range $x\in [0, 1]$. Here, the range for numerical integration is $x\in [10^{-6}, 0.99]$, which gives rise to a technical issue related to normalization. As previously introduced, the nonnegativity of KL divergence inherently requires that both distributions be individually normalized to unity. Otherwise, the calculated KL divergence could yield negative values, which is very counterintuitive for a measure representing any sort of entropy. Hence we introduce two normalization factors $z_p$ and $z_q$, and the normalized distributions are
\begin{eqnarray}\label{DKLnor}
  &&\left\{
  \begin{aligned}
  &\left. p^A(x) \right|_{\textrm{nor}} = \frac{1}{\int_{x_{\textrm{min}}}^{x_{\textrm{max}}}p^A(x) dx} \, p^A(x) = z_p \, p^A(x) \,,\\
  &\left. q^d(x) \right|_{\textrm{nor}} = \frac{1}{\int_{x_{\textrm{min}}}^{x_{\textrm{max}}}q^d(x) dx} \, q^d(x) = z_q \, q^d(x) \,.
  \end{aligned}
  \right.
\end{eqnarray}

The KL divergence between the structure functions of nucleus A and deuteron is calculated through
\begin{eqnarray}\label{DKLfull}
  D_{\mathrm{KL}}(p^A \| q^d) = \int_{x_{\textrm{min}}}^{x_{\textrm{max}}} \left. p^A(x) \right|_{\textrm{nor}} \ln\frac{\left. p^A(x) \right|_{\textrm{nor}}}{\left. q^d(x) \right|_{\textrm{nor}}} \, dx \,.
\end{eqnarray}

Here, $p^A(x)$ denotes the nuclear structure function constructed from the EPPS21 global analysis, and $q^d(x)$ represents the deuteron reference structure function built from the CT18A free PDFs. For the full-$x$ region, more  details are collected in Appendix \ref{AppA}.

While the full-$x$ KL divergence provides a useful global measure of the nuclear modification, the central goal of this work is to determine the shape of the structure function within the intermediate-$x$ EMC region. We will focus on this interval in what follows.
%\red{We now turn to the intermediate-$x$ range, where the nucleon structure function is of particular interest because it is closely related to the EMC effect.} 
In the EMC region, the ratio monotonically decreases from a  maximum to a minimum. Within the EPPS21 set, this region spans $x\in [0.25, 0.65]$. We have also computed the KL divergence within this region using the same method described above, except that the integration limits in Eqs.\,(\ref{DKLnor}) and (\ref{DKLfull}) have been changed from $[10^{-6}, 0.99]$ to $[0.25, 0.65]$. More details about these resulting KL divergences for different nuclei are collected in Appendix \ref{AppA}.

These results essentially quantify the findings which have already been known. More importantly, we are interested in whether the KL divergence can provide new insights into nucleon structure, especially in the EMC region. Prior to the discovery of the EMC effect, it was commonly believed that the ratio $(F_2^A(x)/A)/(F_2^d(x)/2)$ would be unity in the intermediate-$x$ region, implying a corresponding KL divergence of zero. However, the EMC effect revealed that this ratio exhibits a deviation from unity, showing a monotonic decrease in the approximate $x$-range of 0.25 to 0.65.

For a given nucleus A, if we fix the two endpoints at $x = 0.25$ and $x = 0.65$, respectively, a question arises: can we establish a criterion to determine the shape of the structure function within the EMC region? This seems difficult since we are trying to construct this function with insufficient data---neither perturbative QCD nor lattice QCD can be applied on this issue. Interestingly, the problem encountered here bears a striking resemblance to the famous brachistochrone problem proposed by Bernoulli in 1696 \cite{brachistochronecurve}. Both involve finding the extremum of a functional under fixed boundary conditions: the brachistochrone problem seeks the minimal time of descent, while our task requires establishing an extremum principle for a physical observable.

The physical observable we choose is the KL divergence utilized above,
\begin{eqnarray}\label{DKLEMC}
  D_{\mathrm{KL}}^{\textrm{EMC}}(\hat  p^A \| q^d) = \int_{0.25}^{0.65} \left. \hat p^A(x) \right|_{\textrm{nor}} \ln\frac{\left. \hat p^A(x) \right|_{\textrm{nor}}}{\left. q^d(x) \right|_{\textrm{nor}}} \, dx \,.
\end{eqnarray}
In the above equation, the $\hat p^A(x)$  represents the structure function we aim to determine, and $q^d(x)$ denotes the deuteron reference distribution introduced above. We assume that, with the endpoints fixed, the actual structure function of the nucleus tends to minimize the KL divergence in Eq.\,(\ref{DKLEMC}), which we refer to as the ``minimum relative entropy'' hypothesis. It is important to emphasize that this hypothesis serves a role analogous to the shortest time principle in the brachistochrone problem. With this hypothesis, the structure function can be computed and subsequently compared with the results from global fitting analyses.  First, we adopt a polynomial form of parametrization for $\hat p^A(x)$
\begin{eqnarray}\label{polypara}
  \hat p^A_{\textrm{poly}}(x) = \big( a_1 x^3+b_1 x^2+c_1 x+d_1 \big) \, q^d(x) \,.
\end{eqnarray}
Using the EPPS21 global fits, we can derive constraints at the endpoints, which determine two parameters in Eq.\,(\ref{polypara}), while the remaining two are fixed by the minimum relative entropy hypothesis. It is worth noting that the two endpoint values at $x=0.25$ and $x=0.65$ are obtained from the EPPS21 global fits and thus encode direct experimental information about the nuclear modification at the boundaries of the EMC region. An alternative approach would be to employ momentum sum rules as theoretical constraints. However, sum rules are integral constraints over the full $x$-range; they do not directly fix the shape of distributions within the EMC sub-interval. Therefore, we did not adopt this method in the current work.

In Figure \ref{3Dquark}, we provide two-dimensional visualization of the KL divergence as a function of $a_1$ and $b_1$ parameters. \red{\footnote{From the perspective of artificial intelligence, the KL divergence here serves as a loss function defined over the parameter space. Figure \ref{3Dquark} displays the corresponding loss landscape, where the minimum point (red dot) identifies the optimal solution under the imposed endpoint constraints.}} The KL divergence forms a distribution in the parameter space that allows us to pin down its minimum point. The position marked with a red dot indicates the corresponding $a_1$ and $b_1$ with minimum KL divergence. The black dot represents the result of EPPS21 global analyses. These two points are found to be in close proximity within the parameter space. %To further illustrate the effect of deviating from the minimum relative entropy solution, two additional dots, shown in magenta and orange, are chosen to serve as representative examples. 
The structure functions corresponding to all these four dots will be presented in Figure \ref{3xunnormalized}.
\begin{widetext}

\begin{figure}[htbp]
\begin{minipage}{0.49\columnwidth}
\centerline{\includegraphics[width=0.8\columnwidth]{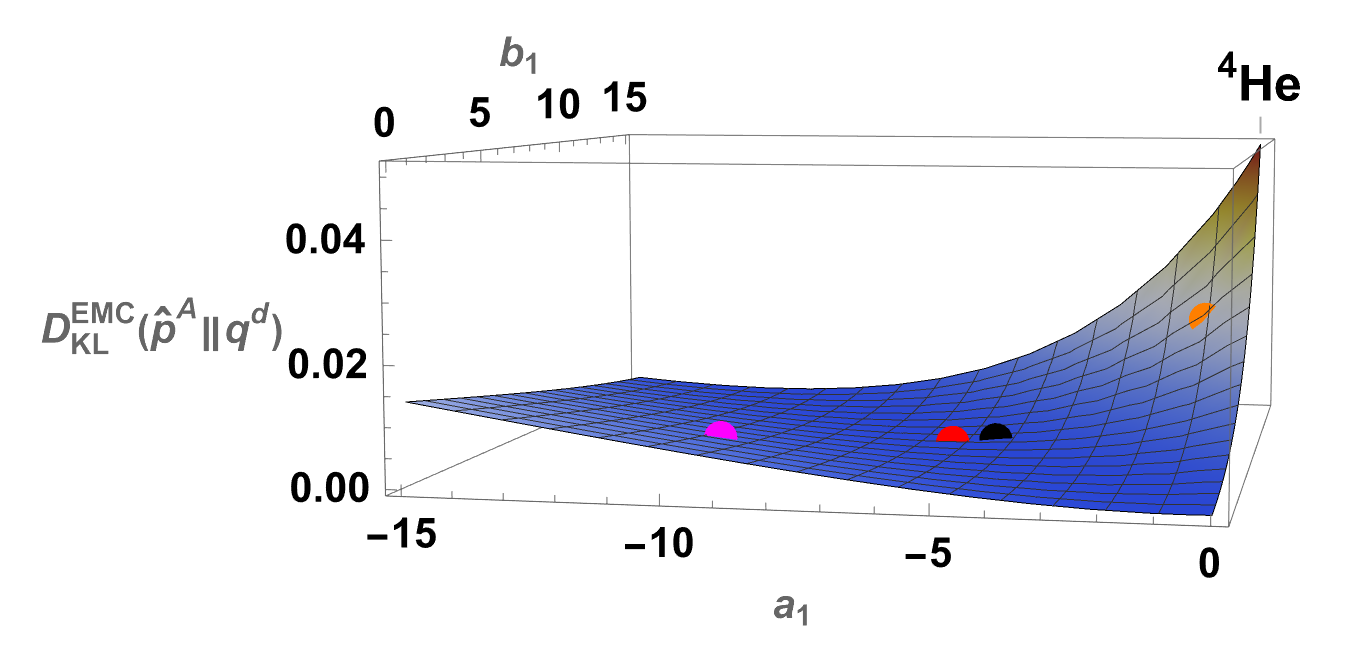}}
\vspace{-2pt}
\centerline{\includegraphics[width=0.8\columnwidth]{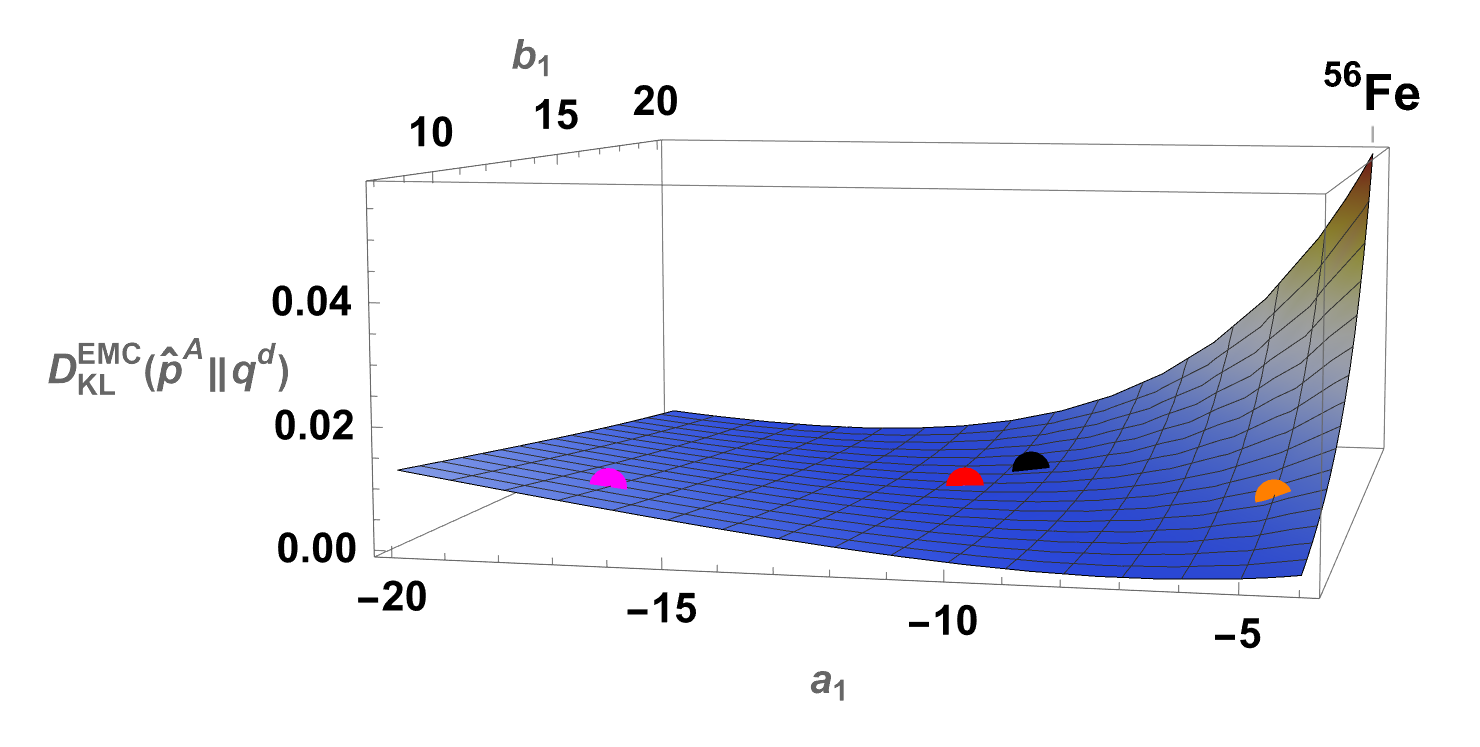}}
\end{minipage}
\begin{minipage}{0.49\columnwidth}
\vspace{0pt}
\centerline{\includegraphics[width=0.8\columnwidth]{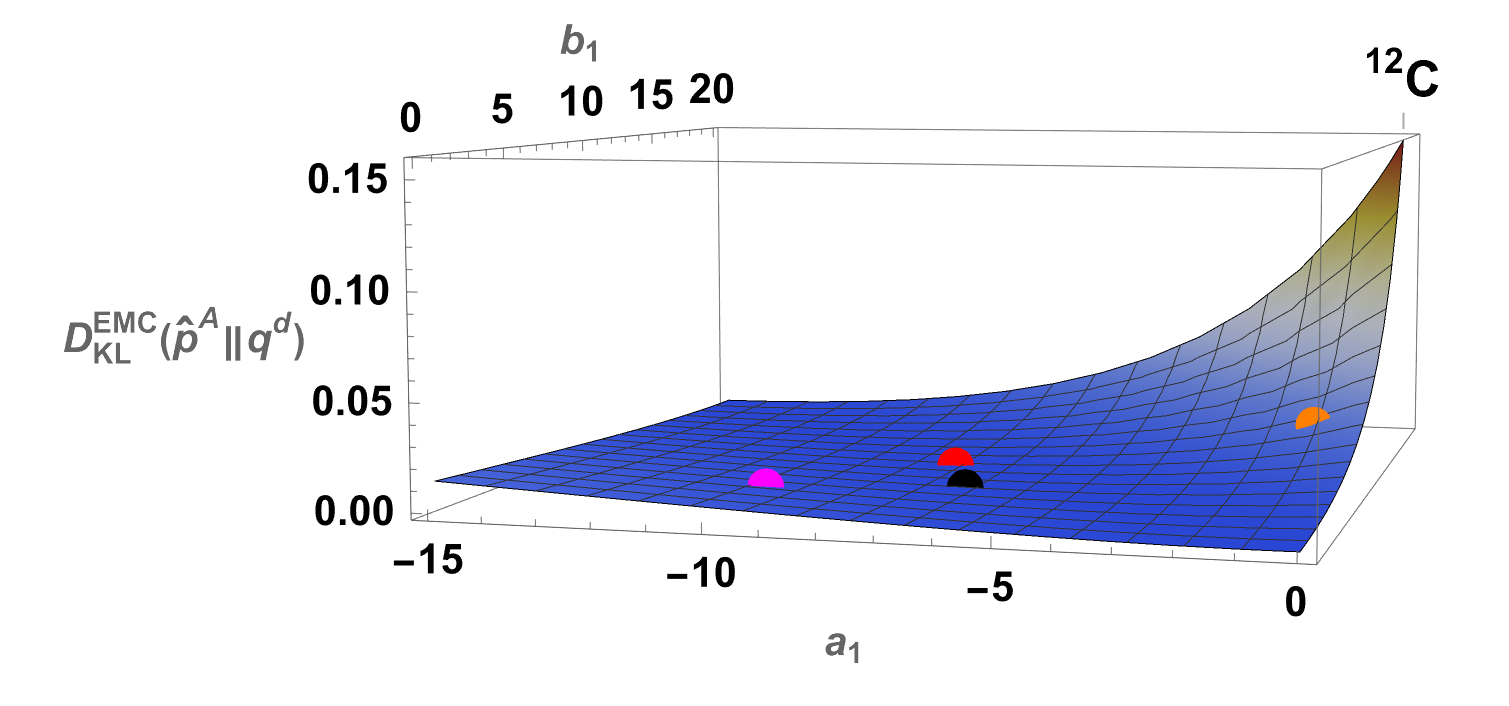}}
\vspace{-2pt}
\centerline{\includegraphics[width=0.8\columnwidth]{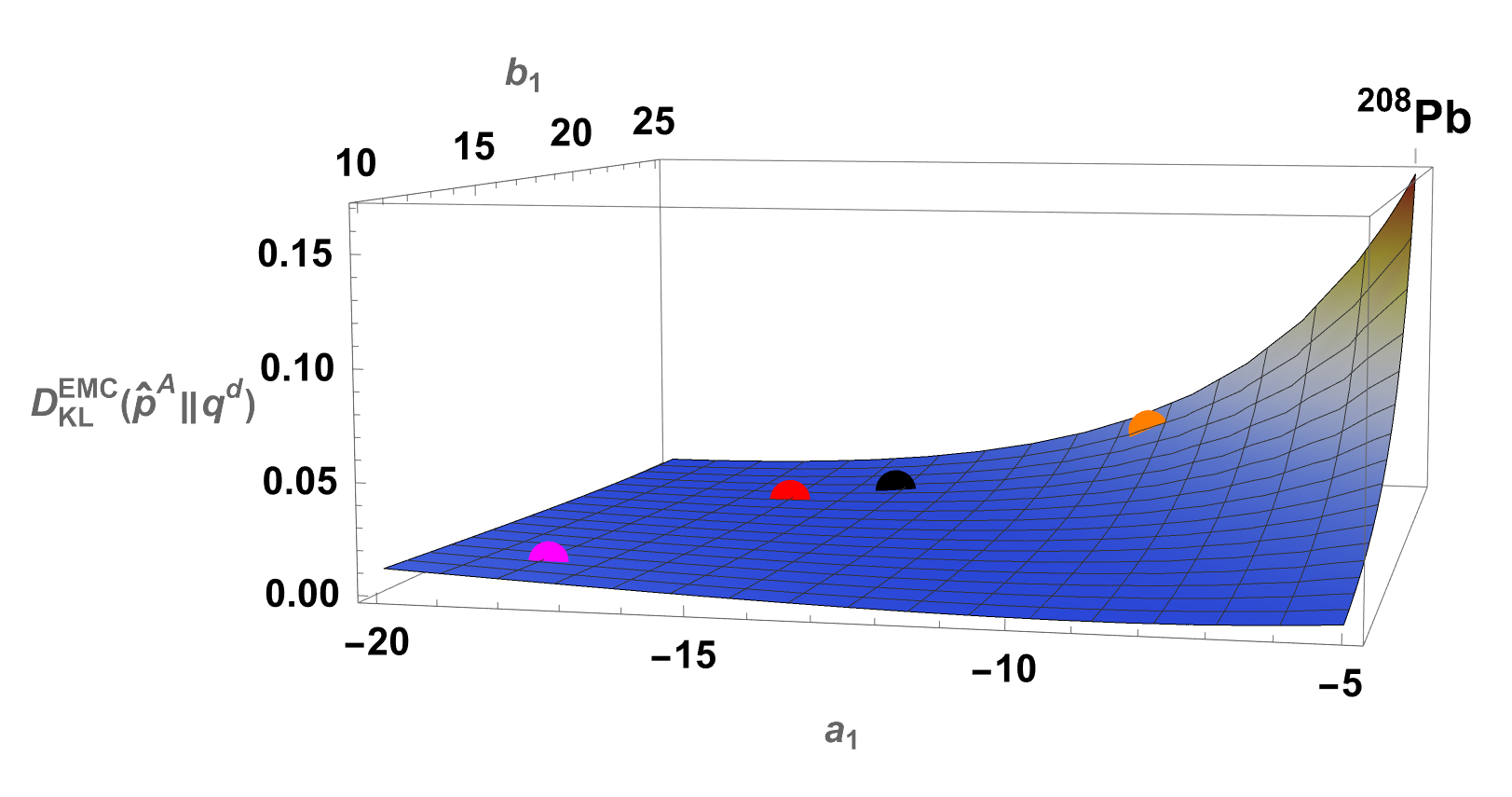}}
\end{minipage}
\caption{
The KL divergence as a function of $a_1$ and $b_1$. The red dot indicates the position with minimum KL divergence for $^{4}$He, $^{12}$C, $^{56}$Fe and $^{208}$Pb, at the momentum transfer $Q^2 = 10\,{\rm GeV}^2$. The orange and magenta dots mark two representative reference solutions, which will be used for comparisons in Figure \ref{3xunnormalized}.}
\label{3Dquark}
\end{figure}

\begin{figure}[htbp]
	\centering
	\begin{minipage}{0.44\columnwidth}
		\centerline{\includegraphics[width=0.8\columnwidth]{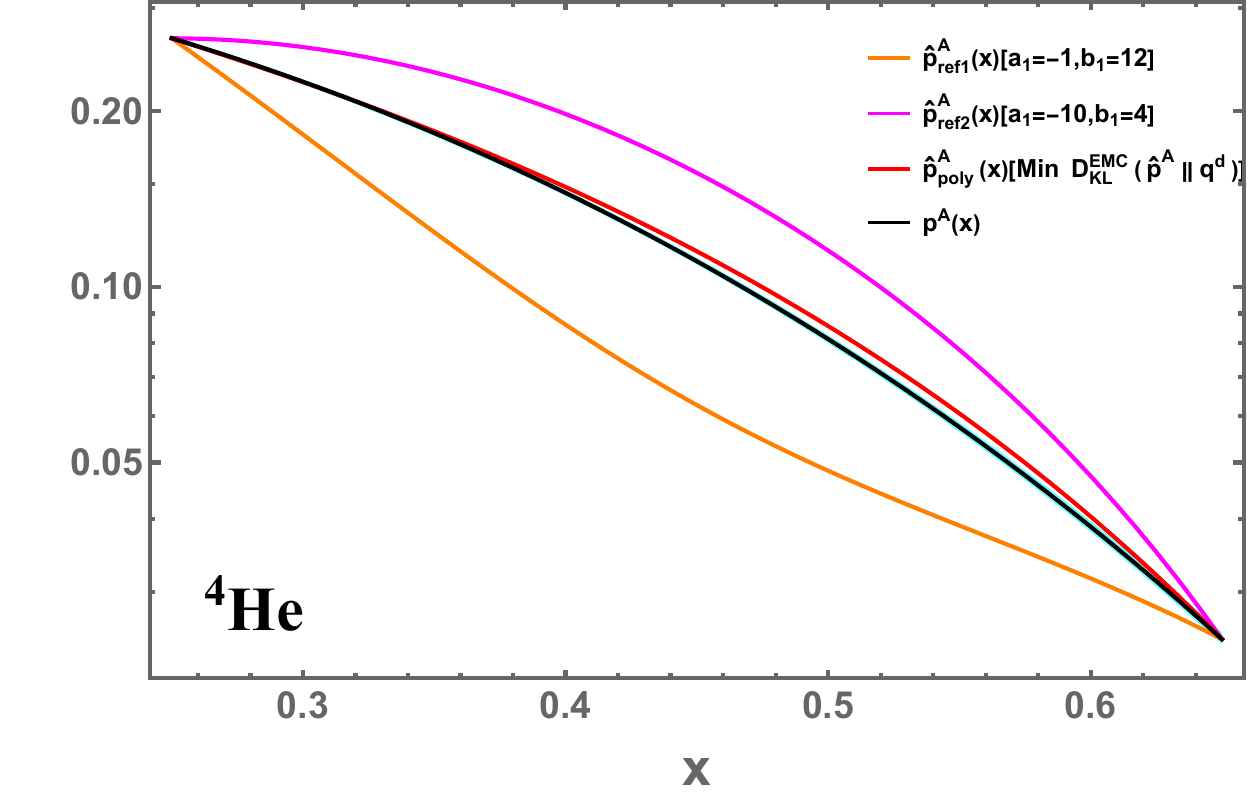}}
		\vspace{10pt}
		\centerline{\includegraphics[width=0.8\columnwidth]{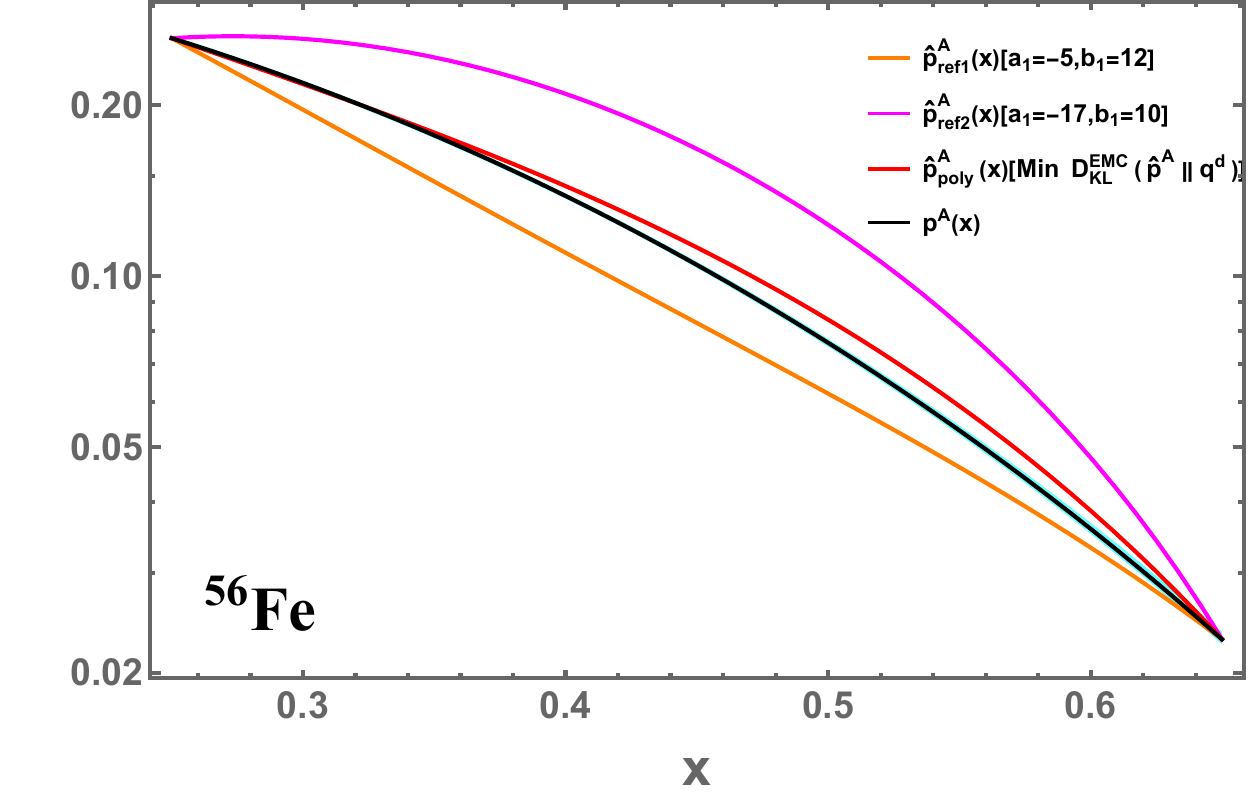}}
	\end{minipage}
	\begin{minipage}{0.44\columnwidth}
		\vspace{0pt}
		\centerline{\includegraphics[width=0.8\columnwidth]{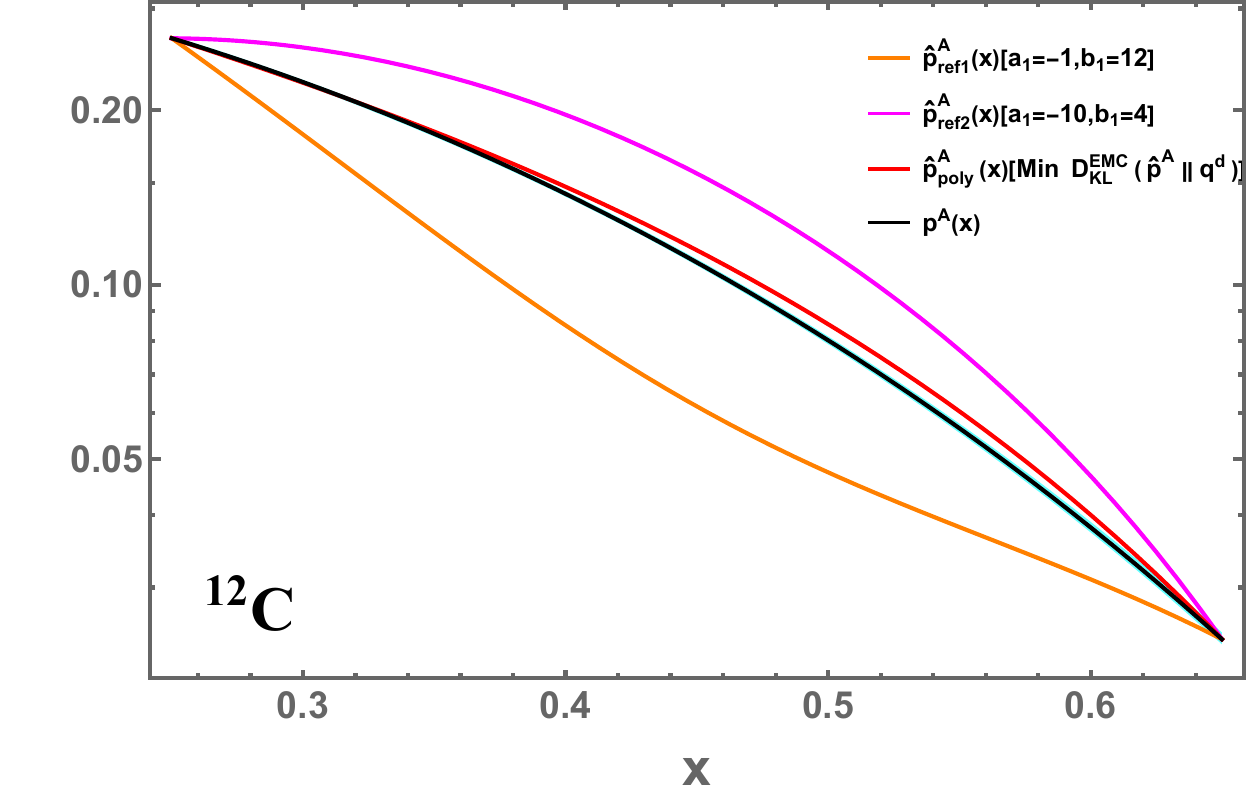}}
		\vspace{10pt}
		\centerline{\includegraphics[width=0.8\columnwidth]{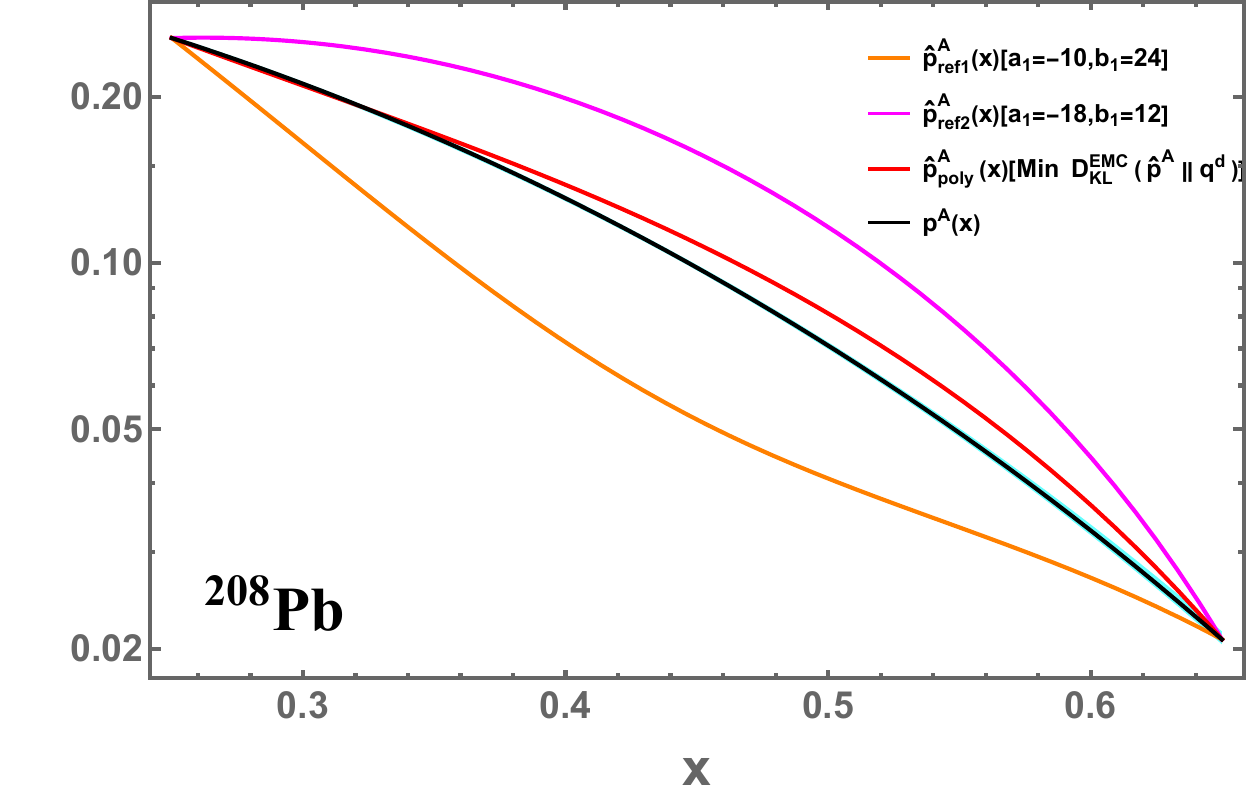}}
	\end{minipage}
	\caption{
    Comparison of the structure functions $\hat{p}^A_{\mathrm{poly}}(x)$ from polynomial parameterization (red line), $p^A(x)$ from EPPS21 (black line), and the reference curves $\hat{p}^A_{\mathrm{ref1}}(x)$ (orange line) and $\hat{p}^A_{\mathrm{ref2}}(x)$ (magenta line).
    }
	\label{3xunnormalized}
\end{figure}

\end{widetext}
We present the explicit results of the polynomial parametrization for $^{4}$He, $^{12}$C, $^{56}$Fe and $^{208}$Pb,
\begin{eqnarray}\label{resultspolypara}
  \hat p^{^{4}\textrm{He}}_{\textrm{poly}}(x) &\!=\!& \big(\! -\! 5.89 \, x^3 \!+\! 7.39 \, x^2 \!-\! 3.00 \, x \!+\! 1.38 \big) \, q^d(x) ,\nn\\
  \hat p^{^{12}\textrm{C}}_{\textrm{poly}}(x) &\!=\!& \big(\! -\! 7.60 \, x^3 \!+\! 9.53 \, x^2 \!-\! 3.86 \, x \!+\! 1.48 \big) \, q^d(x) ,\nn\\
  \hat p^{^{56}\textrm{Fe}}_{\textrm{poly}}(x) &\!=\!& \big(\! -\! 11.51 \, x^3 \!+\! 14.43 \, x^2 \!-\! 5.85 \, x \!+\! 1.72 \big) \, q^d(x) ,\nn\\
  \hat p^{^{208}\textrm{Pb}}_{\textrm{poly}}(x) &\!=\!& \big(\! -\! 16.23 \, x^3 \!+\! 20.32 \, x^2 \!-\! 8.22 \, x \!+\! 1.99 \big) \, q^d(x) .\nn\\
\end{eqnarray}
Appendix \ref{AppB} lists the four parameters ($a_1$, $b_1$, $c_1$ and $d_1$) in Eq.\,(\ref{polypara}).

With the results in Eq.\,(\ref{resultspolypara}), we can plot the structure functions given by the minimum relative entropy hypothesis. Figure \ref{3xunnormalized} shows a comparison between them and the global fitting results from EPPS21 set. Additionally, for reference, we also plot two curves which yield larger KL divergences, corresponding to the orange and magenta dots in Figure \ref{3Dquark}, respectively. From Figure \ref{3xunnormalized}, it can be observed that the structure functions given by Eq.\,(\ref{resultspolypara}) are rather close to the global analysis. This indicates the minimum relative entropy hypothesis can yield reasonable nuclear structure functions, consistent with the results of QCD global analysis.

In addition, we have also examined the dependence of the results on the parametrization form. In Eq.\,(\ref{polypara}), we adopted a polynomial form and expanded it to the cubic term $x^3$. We have also extended the expansion to the quartic term $x^4$, and the resulting $\hat p^A_{\textrm{poly}}(x)$ remained identical to the cubic case. Therefore, we have not shown them for the sake of brevity. Moreover, the widely used canonical parameterization form with exponential and power-law endpoint behaviour was also tested,
\begin{eqnarray}\label{canpara}
	\hat p^A_{\textrm{can}}(x) = \big( a_2 x^{b_2}(1- x)^{c_2}e^{d_2 x} \big) \, q^d(x) \,.
\end{eqnarray}
In line with the previous method, these parameters can be determined by the minimum relative entropy hypothesis,
\begin{eqnarray}\label{resultscanpara}
	\hat p^{^{4}\textrm{He}}_{\textrm{can}}(x) &\!=\!& \big( 0.20 \, x^{-0.69} (1-x)^{1.39} e^{4.13x} \big) \, q^d(x) ,\nn\\
	\hat p^{^{12}\textrm{C}}_{\textrm{can}}(x) &\!=\!& \big( 0.13 \, x^{-0.90} (1-x)^{1.81} e^{5.37x} \big) \, q^d(x) ,\nn\\
	\hat p^{^{56}\textrm{Fe}}_{\textrm{can}}(x) &\!=\!& \big( 0.039 \, x^{-1.40} (1-x)^{2.84} e^{8.43x} \big) \, q^d(x) ,\nn\\
	\hat p^{^{208}\textrm{Pb}}_{\textrm{can}}(x) &\!=\!& \big( 0.008 \, x^{-2.09} (1-x)^{4.25} e^{12.57x} \big) \, q^d(x) .\nn\\
\end{eqnarray}
Appendix \ref{AppB} collects the values of the four parameters ($a_2$, $b_2$, $c_2$ and $d_2$) in Eq.\,(\ref{canpara}).

We now summarize the key findings obtained so far. In Table~\ref{3xKL}, different KL divergences are compiled. The second column lists the KL divergence calculated using the results from the EPPS21 set, while the third and fourth column present the calculated minimum KL divergence from Eq.\,(\ref{resultspolypara}) and Eq.\,(\ref{resultscanpara}), respectively. As one can see, the obtained minimum KL divergences given by the polynomial and canonical parametrization forms show remarkable consistency with each other. This agreement becomes more visually discernible when displaying the structure functions together. Figure \ref{expunnormalized} shows these two kinds of structure functions obtained from the polynomial and canonical parameterizations, as well as the global fitting results from EPPS21 set. Visually indistinguishable differences are observed between the structure functions obtained from these two parameterizations.

It is instructive to examine how the parameters determined by the minimum relative entropy hypothesis depend on the nuclear species. In the polynomial parametrization, the leading cubic coefficient $a_1$ increases in magnitude monotonically from $-5.89$ ($^4$He) to $-16.23$ ($^{208}$Pb), reflecting the progressively stronger nuclear modification for heavier nuclei. Correspondingly, the minimum KL divergence $D_{\mathrm{KL}}^{\mathrm{EMC}}$ generally increases with the nuclear mass number $A$, consistent with the expectation that larger nuclei exhibit more pronounced deviations from the free nucleon baseline. A notable exception is $^{9}$Be, whose KL divergence exceeds that of $^{12}$C. This non-monotonic behavior is consistent with findings in studies of nucleon short-range correlations, and may be attributed to the unique cluster structure of beryllium.

%\red{Notably, although the coefficients in the polynomial and canonical parametrizations vary systematically with nuclear size, the more robust nuclear dependence is reflected in the resulting structure functions and in the KL divergences. In particular, the KL divergence generally increases with nuclear size, while the non-monotonic behavior between $^{9}\textrm{Be}$ and $^{12}\textrm{C}$ is consistent with the special role of $^{9}\textrm{Be}$ discussed in Appendix \ref{AppA}.}

These results demonstrate the robustness of the calculated minimum KL divergence across different parametrization forms. When the two endpoints in the intermediate-$x$ region are fixed, both the value of the minimum KL divergence and the shape of the nuclear structure function are almost exclusively determined by the minimum relative entropy hypothesis. It should be emphasized that, although we adopt two commonly used parametrizations of the structure function, the values of their parameters are not determined by fitting. Instead, they are obtained by applying the minimum relative entropy hypothesis, which selects the parameters that minimize the relative entropy.

\begin{widetext}

\begin{table}[H]
		\centering
		\renewcommand{\arraystretch}{1.5}
		\caption{The comparisons between the calculated KL divergence from EPPS21 and the calculated minimum KL divergences  with the polynomial as well as canonical parameterizations in intermediate-$x$ region $[0.25,0.65]$ for different nuclei.}\label{3xKL}
		\begin{tabular}{c| c c c}
			\hline\hline
			~~~Nucleus~~~ &   
			~~$D_{\mathrm{KL}}^{\textrm{EMC}}( p^A \| q^d)$~~&
			~~$D_{\mathrm{KL}}^{\textrm{EMC}}(\hat p^A_{\textrm{poly}} \| q^d)$~~ & ~~$D_{\mathrm{KL}}^{\textrm{EMC}}(\hat p^A_{\textrm{can}} \| q^d)$~~ \\
			\hline
			$^{3}$He       & $0.46\times 10^{-4}$ &$1.3\times 10^{-5}$	& $1.2\times 10^{-5}$     \\
			$^{4}$He       & $2.32\times 10^{-4}$ &$3.0\times 10^{-5}$	& $2.9\times 10^{-5}$          \\
			$^{9}$Be       & $4.75\times 10^{-4}$ &$7.0\times 10^{-5}$	& $6.9\times 10^{-5}$    	  \\
			$^{12}$C       & $3.70\times 10^{-4}$ &$5.0\times 10^{-5}$	& $5.0\times 10^{-5}$       \\
			$^{56}$Fe      & $8.17\times 10^{-4}$ &$1.21\times 10^{-4}$	& $1.21\times 10^{-4}$       \\
			$^{197}$Au     & $16.19\times 10^{-4}$ &$2.52\times 10^{-4}$	& $2.58\times 10^{-4}$      \\
			$^{208}$Pb     & $16.85\times 10^{-4}$ &$2.63\times 10^{-4}$	& $2.70\times 10^{-4}$       \\
			\hline\hline
		\end{tabular}
	\end{table}

\begin{figure}[htbp]
\begin{minipage}{0.44\columnwidth}
\centerline{\includegraphics[width=0.8\columnwidth]{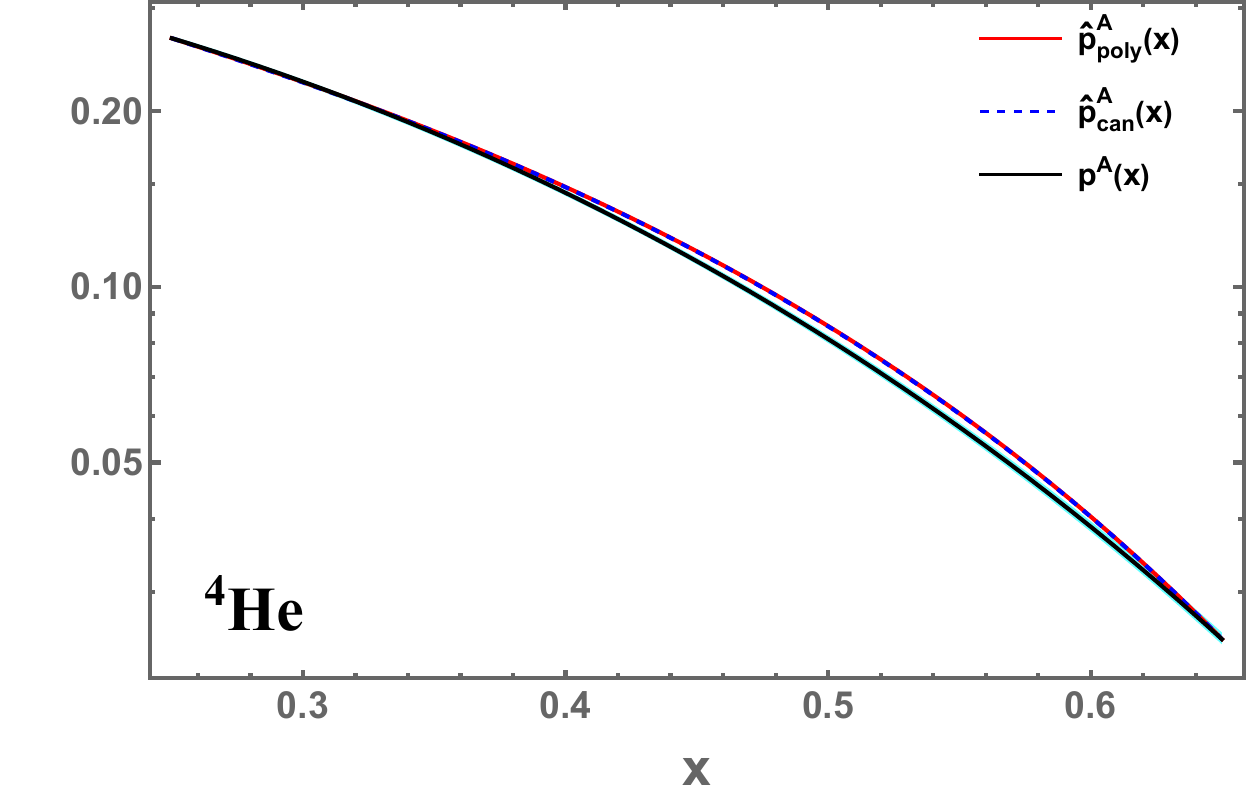}}
\vspace{10pt}
\centerline{\includegraphics[width=0.8\columnwidth]{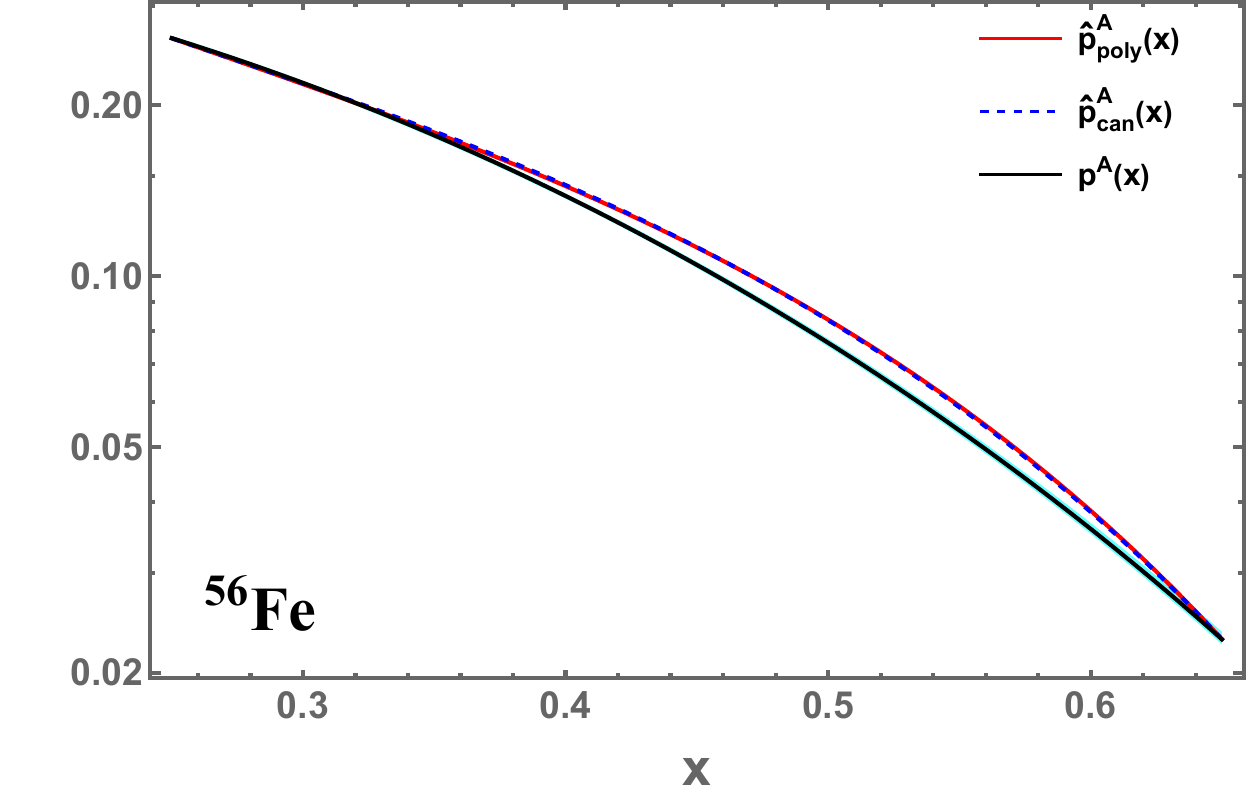}}
\end{minipage}
\begin{minipage}{0.44\columnwidth}
\vspace{0pt}
\centerline{\includegraphics[width=0.8\columnwidth]{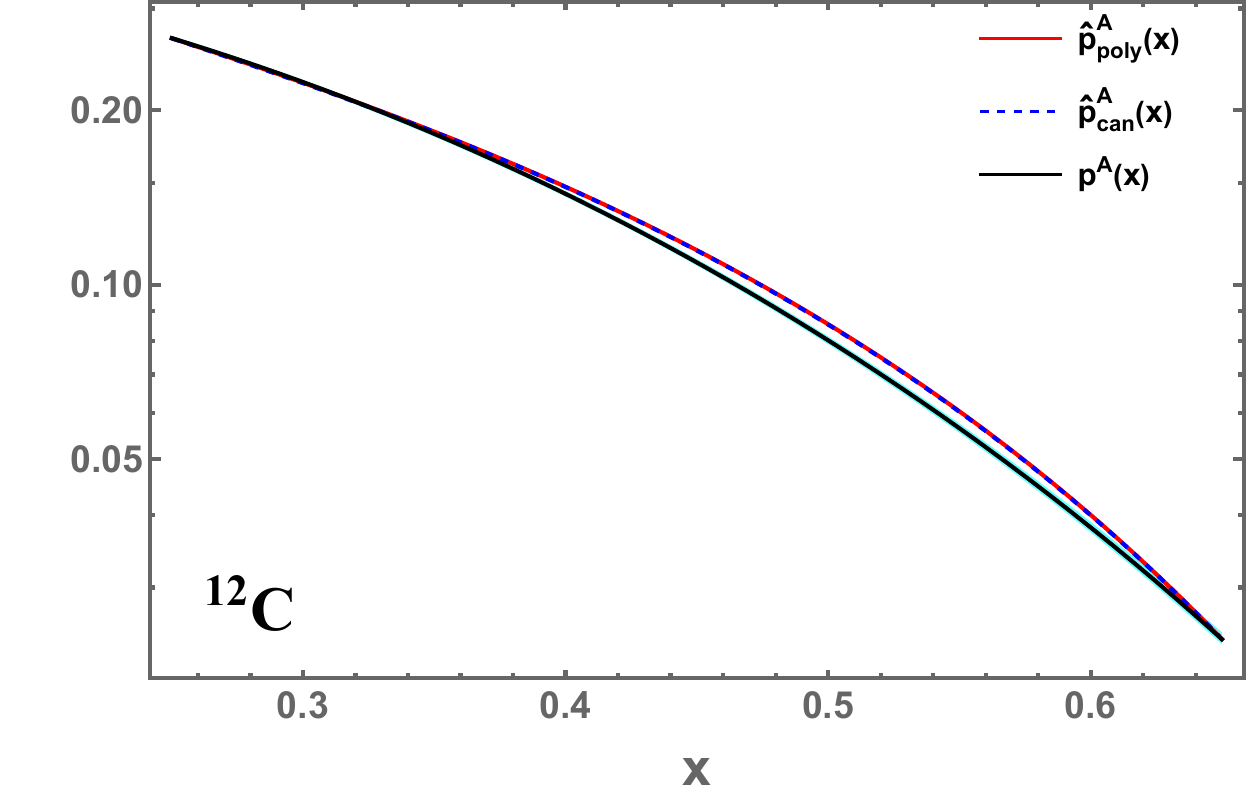}}
\vspace{10pt}
\centerline{\includegraphics[width=0.8\columnwidth]{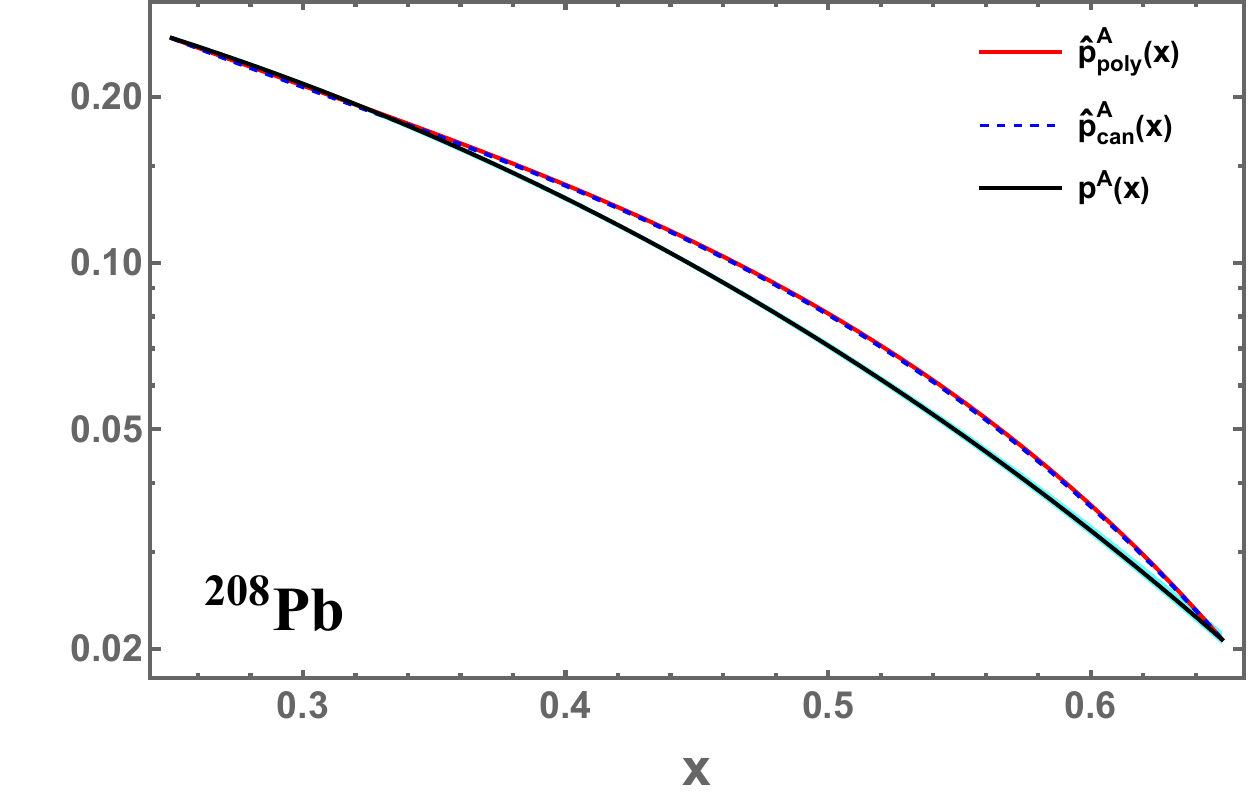}}
\end{minipage}
\caption{Comparison of the structure functions $\hat{p}^A_{\mathrm{poly}}(x)$ from polynomial parameterization  (red line), $\hat{p}^A_{\mathrm{can}}(x)$ from canonical parametrization  (blue dashed line), and $p_A(x)$ from EPPS21 (black line).}
\label{expunnormalized}
\end{figure}

\end{widetext}

To quantify the perceived separation between various structure functions and the global fitting results given by EPPS21 in Figures \ref{3xunnormalized} and \ref{expunnormalized}, we make use of the $L^2$ norm which is defined as
\begin{eqnarray}
  N(\hat p^A_{\textrm{poly}}) = \sqrt{\int_{0.25}^{0.65}\Big( \hat p^A_{\textrm{poly}}(x)-p^A(x) \Big)^2 dx} \,,
\end{eqnarray}
here $p^A(x)$ is the structure function from EPPS21. The results are listed in the second column of Table \ref{3xnorm}. It is evident that the norms calculated from $\hat p^A_{\textrm{poly}}(x)$ are much less than the reference curves $ \hat p^A_{\textrm{ref1}}(x)$ and $\hat p^A_{\textrm{ref2}}(x)$, indicating the minimum relative entropy hypothesis could serve as a reliable criterion in constructing the quark structure functions of bound nucleon. Moreover, the two different parameterization forms (polynomial and canonical) yield nearly identical norms in Table \ref{3xnorm}.

These findings suggest that the discrepancy between the structure functions derived from the minimum relative entropy hypothesis and those obtained from global fits can be quantified by computing their $L^2$ norms. For the quark structure functions considered in this section, the minimum relative entropy hypothesis produces values close to the global fits, resulting in a small $L^2$ norm (on the order of $10^{-3}$ to $10^{-4}$). In the following section, we adopt this norm-based comparison as a potential  reference for evaluating the quality of global fits to gluon nPDFs.

\begin{widetext}

\begin{table}[H]
	\centering
	\renewcommand{\arraystretch}{1.5}
	\caption{The   norms of structure functions with minimum KL divergences (the second column) and norms of reference structure functions (the third and fourth columns).}\label{3xnorm}
	\begin{tabular}{c| c c c}
		\hline\hline
		~~~Nucleus~~~ &  ~~$N(\hat p^A_{\textrm{poly}})$ or $N(\hat p^A_{\textrm{can}})$~~ & ~~~$N(\hat p^A_{\textrm{ref1}})$~~~ & ~~~$N(\hat p^A_{\textrm{ref2}})$~~~ \\
		\hline
		$^{3}$He         &$6.61 \times 10^{-4}$ & $2.11 \times 10^{-2}$ & $2.33 \times 10^{-2}$   \\
		$^{4}$He        &$1.79 \times 10^{-3}$ & $2.50 \times 10^{-2}$ & $2.25 \times 10^{-2}$       \\
		$^{9}$Be       & $2.38 \times 10^{-3}$ & $4.07 \times 10^{-2}$ & $2.91 \times 10^{-2}$   	  \\
  		$^{12}$C       & $2.20 \times 10^{-3}$ & $2.49 \times 10^{-2}$ & $2.26 \times 10^{-2}$      \\
		$^{56}$Fe      & $3.10 \times 10^{-3}$ & $1.22 \times 10^{-2}$ & $3.03 \times 10^{-2}$      \\
		$^{197}$Au     & $4.11 \times 10^{-3}$ & $1.34 \times 10^{-2}$ & $2.31 \times 10^{-2}$      \\
		$^{208}$Pb     & $4.14 \times 10^{-3}$ & $2.53 \times 10^{-2}$ & $2.89 \times 10^{-2}$      \\
		\hline\hline
	\end{tabular}
\end{table}

\end{widetext}

%%%%%%%%%%%%%%%%%%%
\section{Determining gluon nPDFs through KL divergence}
\label{gluonnPDFs}
%%%%%%%%%%%%%%%%%%%

Nucleons are composed of constituent quarks and gluons; therefore, investigating the distribution of gluons, the mediators of the strong interaction, within nuclei is of fundamental importance. However, our understanding of gluon nPDFs remains markedly poorer than that of quark nPDFs. This is reflected in the significant discrepancies among gluon nPDFs extracted by different global fitting groups, especially in the intermediate-$x$ region. Whether gluons also exhibit the  EMC effect is an open question under active investigation \cite{Chen:2016bde,Wang:2016mzo,Xu:2019wso,Hatta:2019ocp,Wang:2020uhj,Wang:2021elw,Wang:2024ikx,Yang:2023zmr,Wang:2022kwg,Huang:2021cac,Wang:2024cpx,Hu:2021fxa}. Our results in the last section show that the minimum relative entropy hypothesis gives reasonable quark structure functions. In this section, we will apply this hypothesis independently to gluon distributions.

We focus on the gluon nPDFs in the intermediate-$x$ region, which vary among different global fitting groups. Here, the results from two widely used global analyses, EPPS21\footnote{https://research.hip.fi/qcdtheory/nuclear-pdfs/epps21/} \cite{Eskola:2021nhw} and nNNPDF3.0\footnote{https://nnpdf.mi.infn.it/for-users/nnnpdf3-0/} \cite{AbdulKhalek:2022fyi} are adopted. The KL divergence between the gluon nPDFs of nucleus A and deuteron is defined as
\begin{eqnarray}\label{DKLgluonEMC}
  D_{\mathrm{KL},g}^{\textrm{EMC}}(p^A_g \| q_g^d) \!=\! \int_{x_{\textrm{min}}}^{x_{\textrm{max}}} \left. p_g^A(x) \right|_{\textrm{nor}} \ln\frac{\left. p_g^A(x) \right|_{\textrm{nor}}}{\left. q_g^d(x) \right|_{\textrm{nor}}} \, dx .
\end{eqnarray}
Here $p_g^A(x)$ is taken as the true distribution and $q_g^d(x)$ as the reference distribution, and their values are given by global fitting groups. The limit of integration is set to $x_{\textrm{min}}=0.25$, $x_{\textrm{max}}=0.50$ for EPPS21 and $x_{\textrm{min}}=0.30$, $x_{\textrm{max}}=0.55$ for nNNPDF3.0.
These distributions necessitate normalization analogous to that in Eq.\,(\ref{DKLnor}).
\iffalse
\begin{eqnarray}\label{DKLnorgluon}
  &&\left\{
  \begin{aligned}
  &\left. p_g^A(x) \right|_{\textrm{nor}} = \frac{1}{\int_{x_{\textrm{min}}}^{x_{\textrm{max}}}p_g^A(x) dx} \, p_g^A(x) = z_{p,g} \, p_g^A(x) \,,\\
  &\left. q_g^d(x) \right|_{\textrm{nor}} = \frac{1}{\int_{x_{\textrm{min}}}^{x_{\textrm{max}}}q_g^d(x) dx} \, q_g^d(x) = z_{q,g} \, q_g^d(x) \,.
  \end{aligned}
  \right.
\end{eqnarray}
\fi
The calculated values for the normalization factors $z_{p,g}$ and $z_{q,g}$ for different nuclei are collected in Appendix \ref{AppA}.

Following the minimum relative entropy hypothesis, we determine the shape of the gluon nPDFs $\hat p_g^A(x)$ that minimizes the relative entropy
\begin{eqnarray}\label{DKLEMCgluon}
  D_{\mathrm{KL},g}^{\textrm{EMC}}(\hat p_g^A \| q_g^d) = \int_{x_{\textrm{min}}}^{x_{\textrm{max}}} \!\! \left. \hat p_g^A(x) \right|_{\textrm{nor}} \ln\frac{\left. \hat p_g^A(x) \right|_{\textrm{nor}}}{\left. q_g^d(x) \right|_{\textrm{nor}}} \, dx .
\end{eqnarray}
Similar to the quark case, the polynomial and canonical form of parameterizations are utilized to calculate $\hat p_g^A(x)$,
\begin{eqnarray}\label{gluonpolypara}
  \hat p^A_{g,\textrm{poly}}(x) &=& \big( a_{1}^g x^3+b_{1}^g x^2+c_{1}^g x + d_{1}^g \big) \, q_g^d(x) \,,\\
  \hat p^A_{g,\textrm{can}}(x) &=& \big( a_{2}^g x^{b_{2}^g}(1- x)^{c_{2}^g}e^{d_{2}^g x} \big) \, q_g^d(x) \,.
  \label{gluoncanpara}
\end{eqnarray}

Following the same procedure outlined in the previous section for quarks, we now directly present the results. They are shown separately according to the global fitting dataset employed.

We begin with EPPS21: the explicit results of $^{4}$He, $^{12}$C, $^{56}$Fe, $^{208}$Pb in the polynomial parametrization are
\begin{eqnarray}\label{gluonresultspolypara}
  \hat p^{^{4}\textrm{He}}_{g,\textrm{poly}}(x) &\!=\!& \big(\! -\! 29.62 \, x^3 \!+\! 31.38 \, x^2 \!-\! 10.89 \, x \!+\! 2.22 \big) \, q_g^d(x) ,\nn\\
  \hat p^{^{12}\textrm{C}}_{g,\textrm{poly}}(x) &\!=\!& \big(\! -\! 36.95 \, x^3 \!+\! 39.15 \, x^2 \!-\! 13.58 \, x \!+\! 2.52 \big) \, q_g^d(x) ,\nn\\
  \hat p^{^{56}\textrm{Fe}}_{g,\textrm{poly}}(x) &\!=\!& \big(\! -\! 50.10 \, x^3 \!+\! 53.05 \, x^2 \!-\! 18.40 \, x \!+\! 3.06 \big) \, q_g^d(x) ,\nn\\
  \hat p^{^{208}\textrm{Pb}}_{g,\textrm{poly}}(x) &\!=\!& \big(\! -\! 64.57 \, x^3 \!+\! 68.34 \, x^2 \!-\! 23.69 \, x \!+\! 3.66 \big) \, q_g^d(x) ,\nn\\
\end{eqnarray}
and
\begin{eqnarray}\label{gluonresultscanpara}
  \hat p^{^{4}\textrm{He}}_{g,\textrm{can}}(x) &\!=\!& \big( 0.002 \, x^{-2.56} (1-x)^{8.90} e^{21.19 x} \big) \, q_g^d(x) ,\nn\\
  \hat p^{^{12}\textrm{C}}_{g,\textrm{can}}(x) &\!=\!& \big( 0.0004 \, x^{-3.22} (1-x)^{11.20} e^{26.68 x} \big) \, q_g^d(x) ,\nn\\
  \hat p^{^{56}\textrm{Fe}}_{g,\textrm{can}}(x) &\!=\!& \big( 0.003 \, x^{-2.37} (1-x)^{9.83} e^{21.94 x} \big) \, q_g^d(x) ,\nn\\
  \hat p^{^{208}\textrm{Pb}}_{g,\textrm{can}}(x) &\!=\!& \big( 0.002 \, x^{-2.36} (1-x)^{10.92} e^{23.50 x} \big) \, q_g^d(x) ,
\end{eqnarray}
for the canonical parametrization. Appendix \ref{AppB} collects the calculated parameters ($a_1^g$, $b_1^g$, $c_1^g$ and $d_1^g$) in Eq.\,(\ref{gluonresultspolypara}) and ($a_2^g$, $b_2^g$, $c_2^g$ and $d_2^g$) in Eq.\,(\ref{gluonresultscanpara}) for different nuclei.

\begin{widetext}    

\begin{figure}[H]
\centering
		\begin{minipage}{0.44\columnwidth}
			\centerline{\includegraphics[width=0.8\columnwidth]{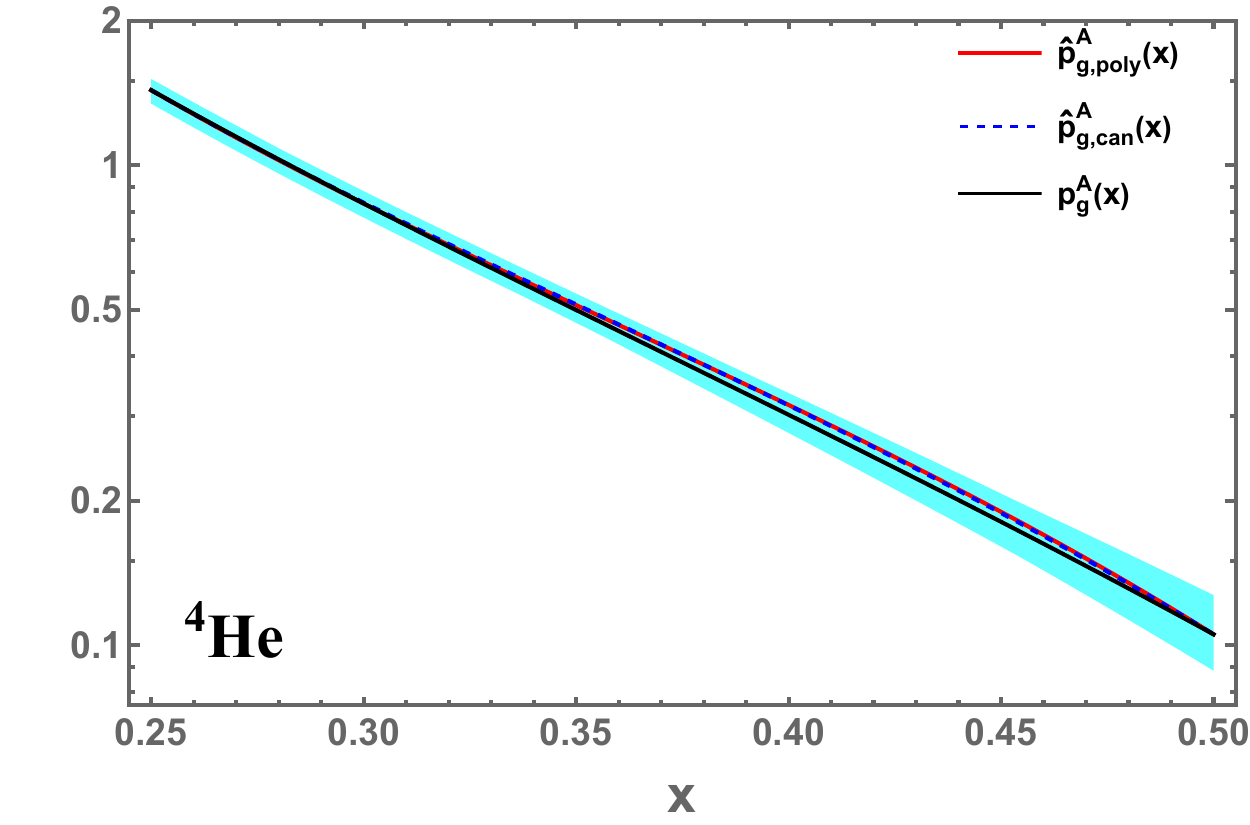}}
			\vspace{10pt}
			\centerline{\includegraphics[width=0.8\columnwidth]{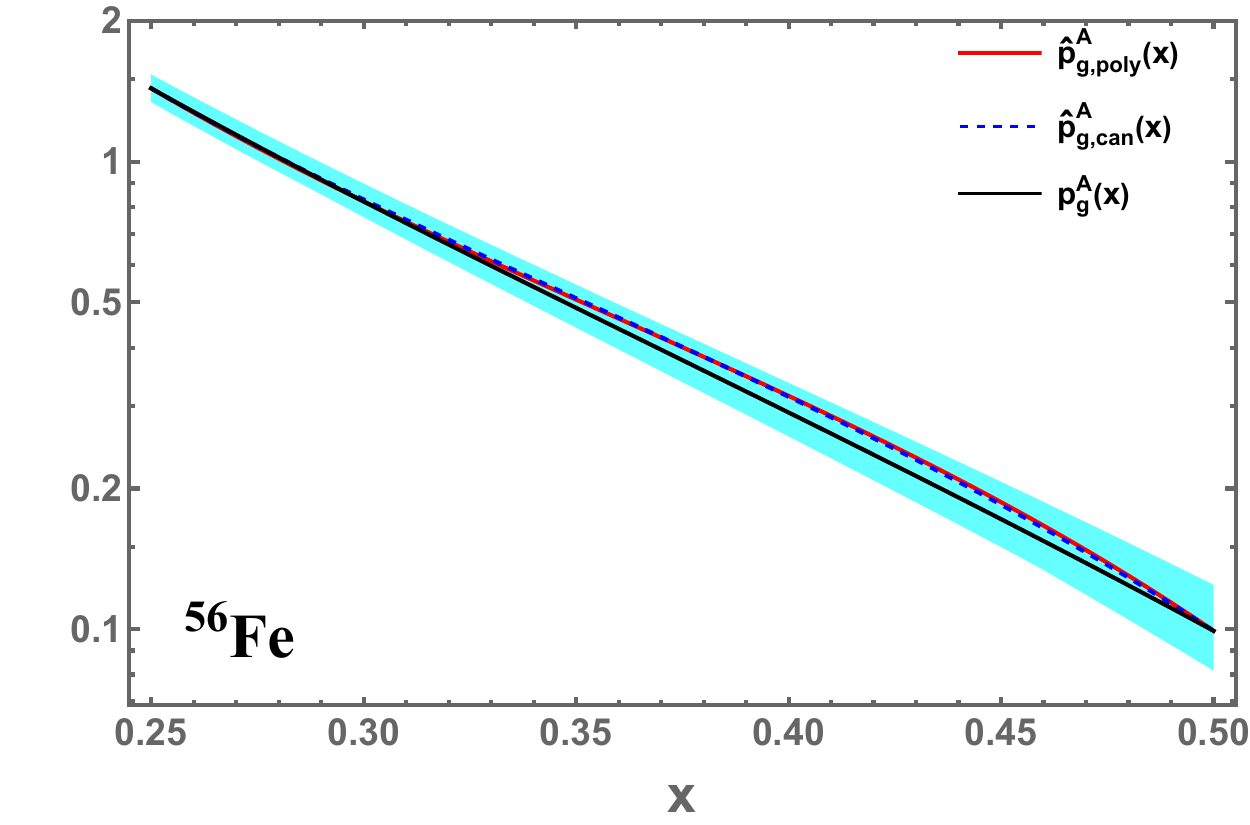}}
		\end{minipage}  
		\begin{minipage}{0.44\columnwidth}
			\vspace{0pt}
			\centerline{\includegraphics[width=0.8\columnwidth]{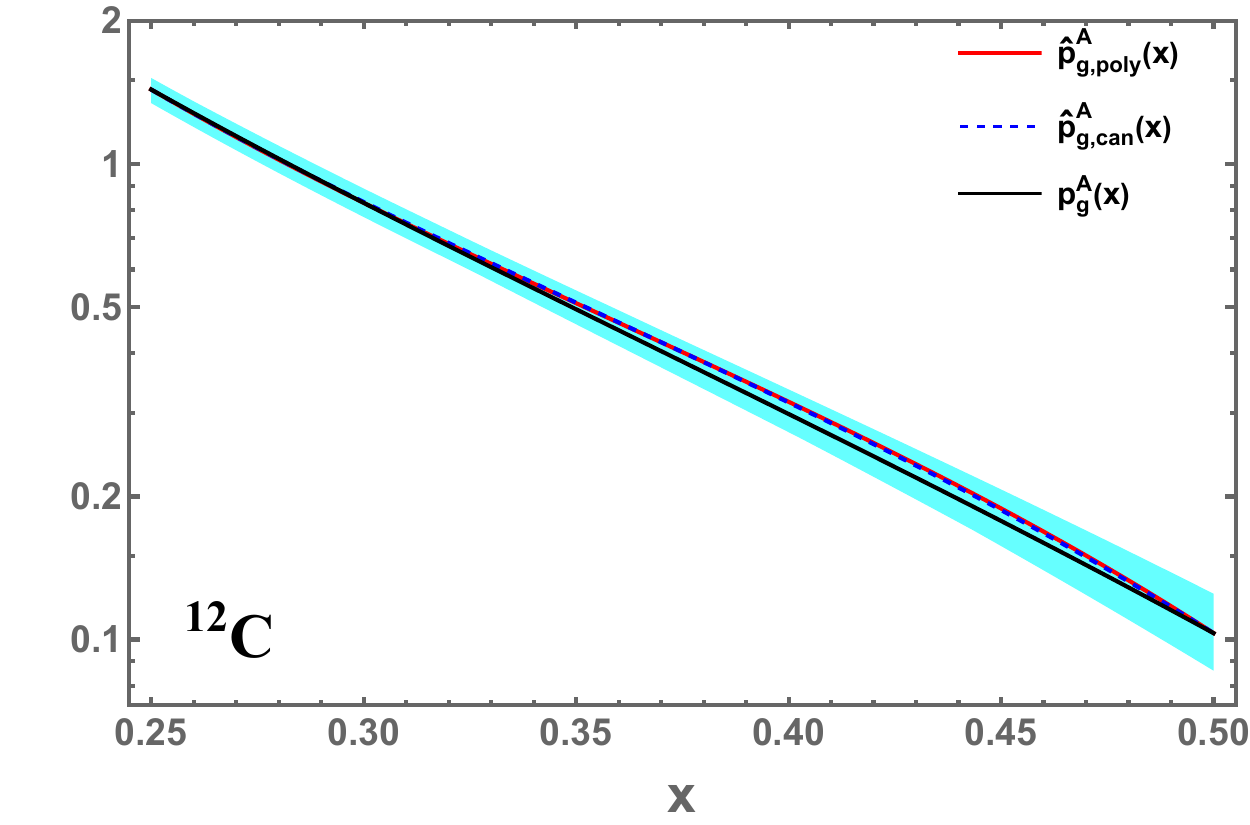}}
			\vspace{10pt}
			\centerline{\includegraphics[width=0.8\columnwidth]{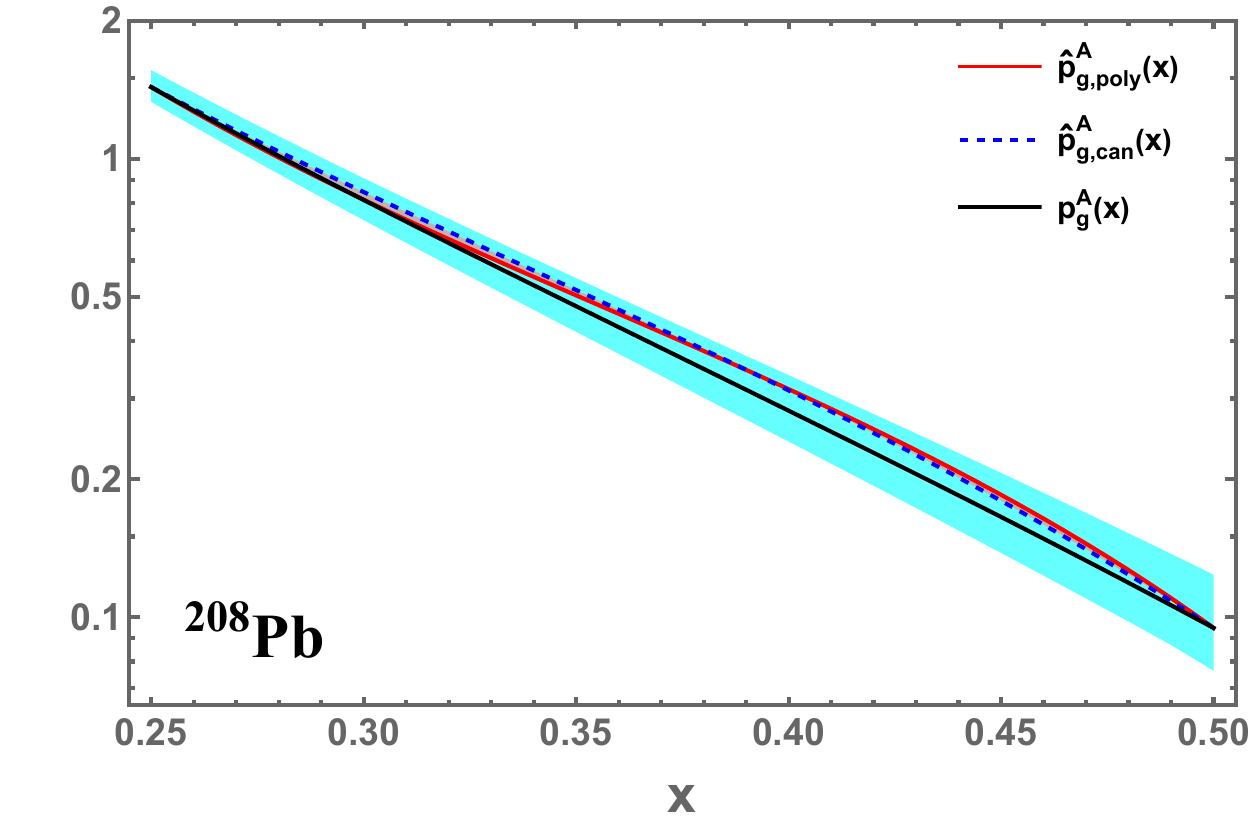}}
		\end{minipage}
		\caption{
        Comparison of the gluon nPDFs $\hat{p}^A_{g,\mathrm{poly}}(x)$ in polynomial parametrization (red line), $\hat{p}^A_{g,\mathrm{can}}(x)$ in canonical parametrization (blue dashed line), and $p^A_g(x)$ from EPPS21 (black line).}
		\label{gluon3xunnormalized1}
	\end{figure}

\end{widetext}

With the results in Eqs.\,(\ref{gluonresultspolypara}) and (\ref{gluonresultscanpara}), we plot the gluon nPDFs given by the minimum relative entropy hypothesis in Figure \ref{gluon3xunnormalized1}. It can be observed that the shapes of gluon nPDFs given by the two parameterizations are highly consistent, with errors between them barely discernible. Moreover, both results agree well with the global fits of experimental data, falling within the error bar of EPPS21 result. It is worth noting that the gluon nPDF has relatively large uncertainties due to our currently poor knowledge of it. Therefore, the minimum relative entropy hypothesis may provide an even more useful complementary perspective for studying gluon nPDFs than for constraining quark structure functions.

Table \ref{gluon3xKL} collects the different KL divergences. The second column lists the KL divergence calculated using the results of gluon nPDFs from EPPS21 set, while the third and fourth columns correspond to the calculated minimum KL divergence according to  Eq.\,(\ref{gluonresultspolypara}) and Eq.\,(\ref{gluonresultscanpara}). As observed in the case of quark structure function, the choice of parametrization does not significantly affect the values of the minimum KL divergence. For the dependence of the determined parameters on the nuclear species, a similar systematic nuclei dependence is observed in the gluon sector: the polynomial coefficient $a_1^g$ ranges from $-29.62$ ($^4$He) to $-64.57$ ($^{208}$Pb), indicating that gluon nuclear modifications also grow with $A$.

%\red{Meanwhile, the parametrization coefficients in the gluon case also vary systematically with nuclear size, while the more robust nuclear dependence is reflected in the resulting distributions and KL divergences.} 

\vspace{-0.015cm}

Similarly, the $L^2$  norm is utilized to quantify the differences between parametrization results and the global fitting results.
\iffalse
\begin{eqnarray}
  N(\hat p^A_{g,\textrm{poly}}) = \sqrt{\int_{x_{\textrm{min}}}^{x_{\textrm{max}}}\Big( \hat p^A_{g,\textrm{poly}}(x)-p_g^A(x) \Big)^2 dx} \,,
\end{eqnarray}
\fi
The norm  values are collected in Table \ref{gluon3xnorm1}. It can be seen that the norms given by the two parameterizations differ very little. This supports the independence of the results with respect to the forms of parameterizations. It can be observed that the $L^2$ norms between the gluon nPDFs derived from the minimum relative entropy hypothesis and those obtained from the EPPS21 global fits range from $1.27\times 10^{-3}$ for $^{3}$He to $1.05\times 10^{-2}$ for $^{208}$Pb, providing a quantitative measure of the discrepancy between the two.

\begin{widetext}

\begin{table}[!htbp]
	\centering
	\renewcommand{\arraystretch}{1.5}
	\caption{
    The KL divergences from EPPS21 and those calculated using polynomial and canonical parameterizations in the intermediate-$x$ region $[0.25,0.50]$. The four-momentum transfer is chosen as $Q^2= 10\,\textrm{GeV}^2$. }\label{gluon3xKL}
	\begin{tabular}{c|c c c}
		\hline\hline
		~~Nucleus~~ & $D_{\mathrm{KL},g}^{\textrm{EMC}}(p^A_g \| q_g^d)$ & $D_{\mathrm{KL},g}^{\textrm{EMC}}(\hat p^A_{g,\textrm{poly}} \| q_g^d)$ & $D_{\mathrm{KL},g}^{\textrm{EMC}}(\hat p^A_{g,\textrm{can}} \| q_g^d)$ \\
		\hline
		$^{3}$He    &  $1.54\times 10^{-5}$   &  $3.34\times 10^{-6}$	& $3.44\times 10^{-6}$  \\
		$^{4}$He    &  $2.31\times 10^{-4}$   &  $4.66\times 10^{-5}$	& $4.86\times 10^{-5}$  \\
		$^{9}$Be    &  $3.24\times 10^{-4}$   &  $6.51\times 10^{-5}$	& $7.19\times 10^{-5}$  \\
		$^{12}$C    &  $3.65\times 10^{-4}$   &  $7.33\times 10^{-5}$	& $7.70\times 10^{-5}$  \\
		$^{56}$Fe   &  $6.91\times 10^{-4}$   &  $1.37\times 10^{-4}$	& $1.98\times 10^{-4}$  \\
		$^{197}$Au  &  $1.17\times 10^{-3}$   &  $2.28\times 10^{-4}$	& $4.21\times 10^{-4}$  \\
		$^{208}$Pb  &  $1.19\times 10^{-3}$   &  $2.33\times 10^{-4}$	& $3.95\times 10^{-4}$  \\
		\hline\hline
	\end{tabular}
\end{table}

\begin{table}[!htbp]
	\centering
	\renewcommand{\arraystretch}{1.5}
	\caption{
    The  norms of gluon nPDFs with minimum KL divergences. }\label{gluon3xnorm1}
	\begin{tabular}{c| c c}
		\hline\hline
		~~~Nucleus~~~ & ~~$N(\hat p^A_{g,\textrm{poly}})$~~ & ~~$N(\hat p^A_{g,\textrm{can}})$~~ \\
		\hline
		$^{3}$He    &$1.27\times 10^{-3}$  &$1.25\times 10^{-3}$ \\
		$^{4}$He    &$4.70\times 10^{-3}$  &$4.58\times 10^{-3}$\\
		$^{9}$Be    &$5.54\times 10^{-3}$  &$5.45\times 10^{-3}$ \\
		$^{12}$C    &$5.87\times 10^{-3}$  &$5.70\times 10^{-3}$ \\
		$^{56}$Fe   &$8.02\times 10^{-3}$  &$9.05\times 10^{-3}$ \\
		$^{197}$Au  &$1.03\times 10^{-2}$  &$1.15\times 10^{-2}$ \\
		$^{208}$Pb  &$1.05\times 10^{-2}$  &$1.28\times 10^{-2}$ \\
		\hline\hline
	\end{tabular}
\end{table}

\end{widetext}

 We turn next to nNNPDF3.0. Applying this set of gluon global fitting data in conjunction with the minimum relative entropy hypothesis, we obtain results for $^{4}$He, $^{12}$C, $^{56}$Fe, and $^{208}$Pb within the polynomial parametrization
\begin{eqnarray}\label{gluonresultspolypara3.0}
  \hat p^{^{4}\textrm{He}}_{g,\textrm{poly}}(x) &\!\!\!=\!\!\!& \big(\! -\! 8.97 \, x^3 \!+\! 10.96 \, x^2 \!-\! 4.41 \, x \!+\! 1.60 \big) \, q_g^d(x) ,\nn\\
  \hat p^{^{12}\textrm{C}}_{g,\textrm{poly}}(x) &\!\!\!=\!\!\!& \big(\! -\! 47.08 \, x^3 \!+\!57.52 \, x^2 \!-\! 23.11 \, x \!+\! 4.13 \big) \, q_g^d(x) ,\nn\\
  \hat p^{^{56}\textrm{Fe}}_{g,\textrm{poly}}(x) &\!\!\!=\!\!\!& \big(\! -\! 216.67 \, x^3 \!+\! 263.53 \, x^2 \!-\! 105.39 \, x \!+\! 15.07 \big) \, q_g^d(x) ,\nn\\
  \hat p^{^{208}\textrm{Pb}}_{g,\textrm{poly}}(x) &\!\!\!=\!\!\!& \big(\! -\! 389.23 \, x^3 \!+\! 470.88 \, x^2 \!-\! 187.19 \, x \!+\! 25.70 \big) \, q_g^d(x) .\nn\\
\end{eqnarray}

The corresponding canonical parametrization is
\begin{eqnarray}\label{gluonresultscanpara3.0}
  \hat p^{^{4}\textrm{He}}_{g,\textrm{can}}(x) &\!=\!& \big( 0.0976 \, x^{-1.03} (1-x)^{2.22} e^{6.33 x} \big) \, q_g^d(x) ,\nn\\
  \hat p^{^{12}\textrm{C}}_{g,\textrm{can}}(x) &\!=\!& \big(0.0014  \, x^{-2.86} (1-x)^{7.16} e^{19.15 x} \big) \, q_g^d(x) ,\nn\\
  \hat p^{^{56}\textrm{Fe}}_{g,\textrm{can}}(x) &\!=\!& \big(0.0025  \, x^{-2.20} (1-x)^{15.47} e^{30.52 x} \big) \, q_g^d(x) ,\nn\\
  \hat p^{^{208}\textrm{Pb}}_{g,\textrm{can}}(x) &\!=\!& \big(0.0002 \, x^{-2.60} (1-x)^{31.26} e^{56.80 x} \big) \, q_g^d(x) .\nn\\
\end{eqnarray}
Appendix \ref{AppB} collects the calculated parameters ($a_1^g$, $b_1^g$, $c_1^g$ and $d_1^g$) in Eq.\,(\ref{gluonresultspolypara3.0}) and ($a_2^g$, $b_2^g$, $c_2^g$ and $d_2^g$) in Eq.\,(\ref{gluonresultscanpara3.0}).

The gluon nPDFs obtained from the minimum relative entropy hypothesis are shown in Fig.\,\ref{gluon-nNNPDF30}. Interestingly, these nPDFs begin to exhibit noticeable deviations from the nNNPDF3.0 results, with the discrepancy becoming particularly pronounced for $^{56}$Fe and $^{208}$Pb. Their corresponding KL divergences are shown in  Table \ref{gluon3xKL30}. This discrepancy is more clearly illustrated by computing the $L^2$ norms between the gluon nPDFs derived from the minimum relative entropy hypothesis and those obtained from the nNNPDF3.0 global fits. These values are presented in Table~\ref{gluon3xnorm130}, where the norms are generally larger than those obtained with EPPS21 previously. For example, the results for $^{56}$Fe, $^{197}$Au, and $^{208}$Pb are approximately four times larger than their EPPS21 counterparts. At the level of central values, the EPPS21 gluon distributions are closer to the minimum relative entropy solutions obtained in the present analysis. However, since nNNPDF3.0 carries significantly larger uncertainties and the endpoint uncertainties have not yet been considered, this comparison based on the minimum relative entropy hypothesis should be regarded as a potential reference rather than a benchmark.

\begin{widetext}
	
	\begin{figure}[H]
    \centering
		\begin{minipage}{0.45\columnwidth}
			\centerline{\includegraphics[width=0.8\columnwidth]{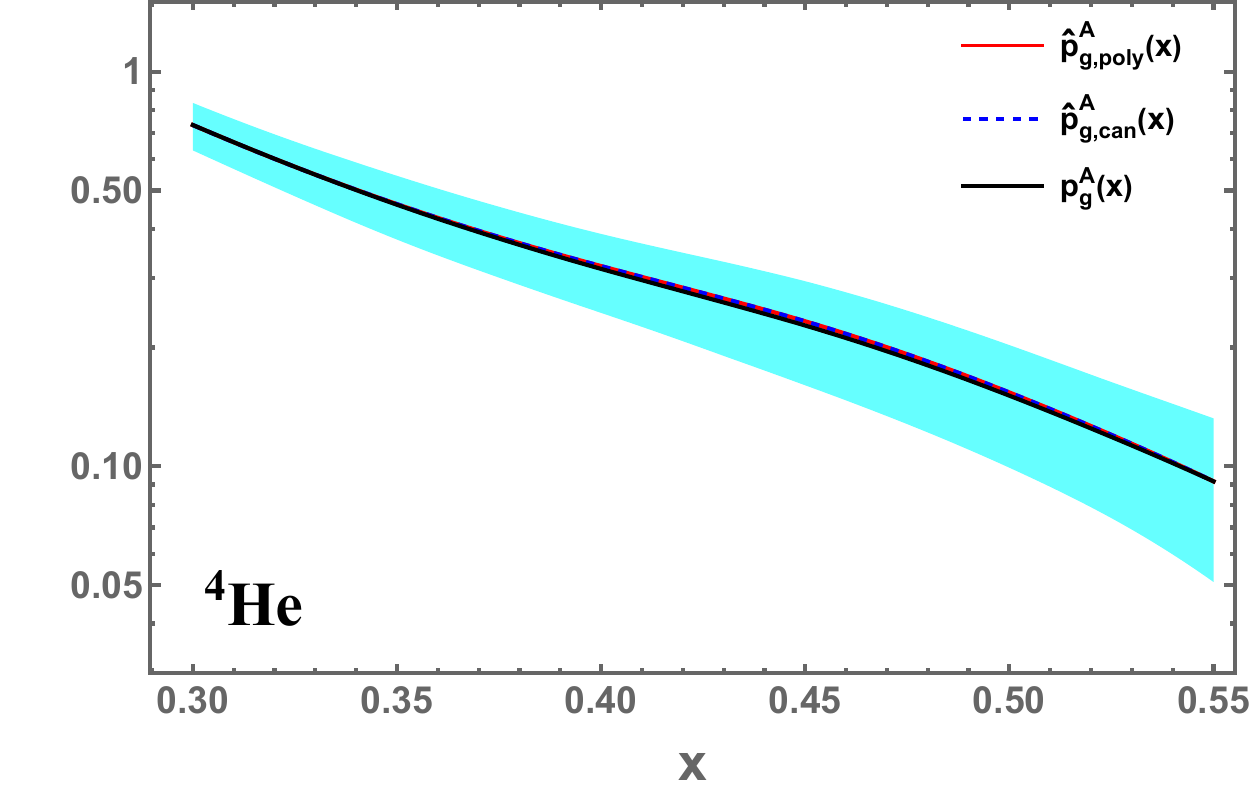}}
			\vspace{10pt}
			\centerline{\includegraphics[width=0.8\columnwidth]{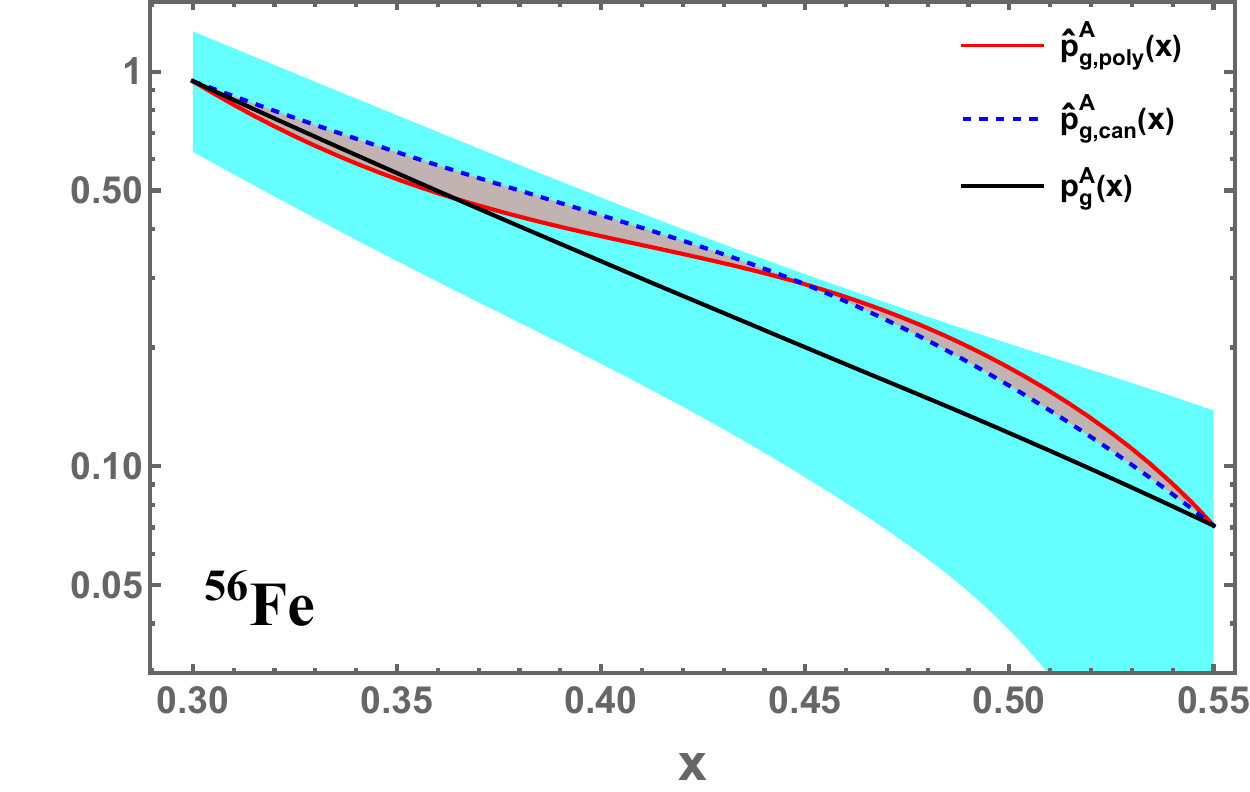}}
		\end{minipage}
		\begin{minipage}{0.45\columnwidth}
			\vspace{0pt}
			\centerline{\includegraphics[width=0.8\columnwidth]{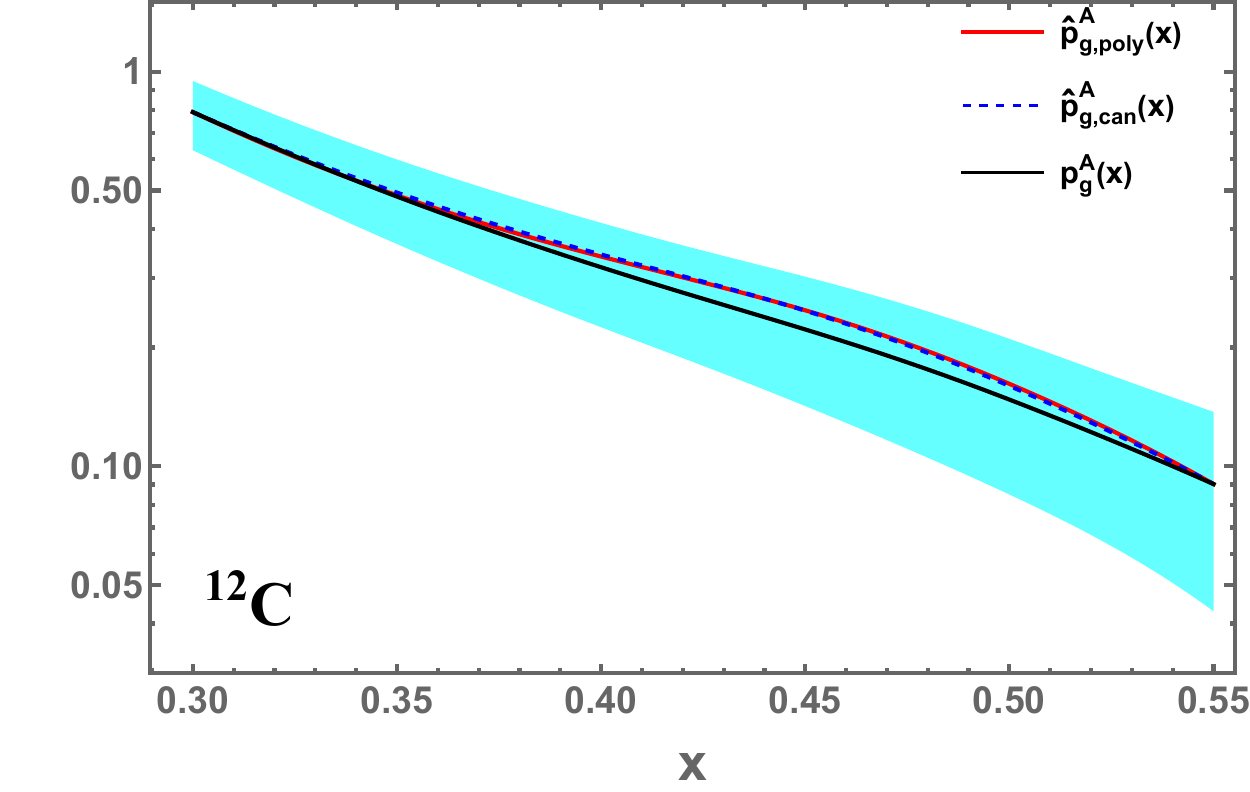}}
			\vspace{10pt}
			\centerline{\includegraphics[width=0.8\columnwidth]{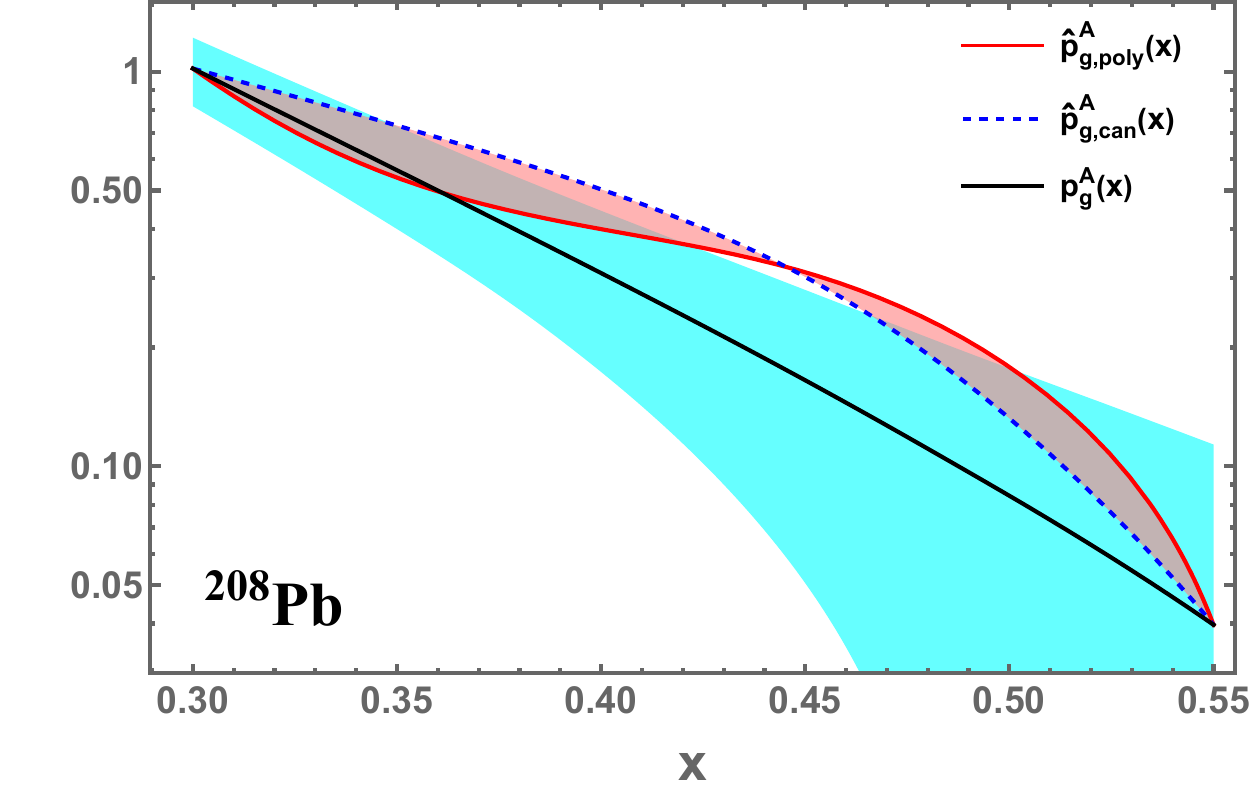}}
		\end{minipage}
		\caption{Same as Figure \ref{gluon3xunnormalized1}, but for nNNPDF3.0. 
        \iffalse
         The obtained gluon nPDFs $\hat p^{A}_{g,\textrm{poly}}(x)$ with minimum KL divergence in polynomial parametrization (red line), the $\hat p^{A}_{g,\textrm{can}}(x)$ with minimum KL divergence in canonical parametrization (blue dashed line), as well as the gluon nPDFs $p_g^A(x)$ given by the global analysis from nNNPDF3.0 (black line).
         \fi}
		\label{gluon-nNNPDF30}
	\end{figure}

\begin{table}[!htbp]
	\centering
	\renewcommand{\arraystretch}{1.5}
	\caption{ Same as Table~\ref{gluon3xKL}, but for nNNPDF3.0 gluon nPDFs in the intermediate-$x$ region $[0.30,0.55]$.
    } \label{gluon3xKL30}
	\begin{tabular}{c|c c c}
		\hline\hline
		~~Nucleus~~ & $D_{\mathrm{KL},g}^{\textrm{EMC}}(p^A_g \| q_g^d)$ & $D_{\mathrm{KL},g}^{\textrm{EMC}}(\hat p^A_{g,\textrm{poly}} \| q_g^d)$ & $D_{\mathrm{KL},g}^{\textrm{EMC}}(\hat p^A_{g,\textrm{can}} \| q_g^d)$ \\
		\hline
		$^{4}$He    &  $4.55\times 10^{-5}$   &  $4.27\times 10^{-6}$	& $4.30\times 10^{-6}$  \\
		$^{9}$Be    &  $5.23\times 10^{-4}$   &  $5.34\times 10^{-5}$	& $6.93\times 10^{-5}$  \\
		$^{12}$C    &  $1.02\times 10^{-3}$   &  $1.07\times 10^{-4}$	& $1.50\times 10^{-4}$  \\
		$^{56}$Fe   &  $1.45\times 10^{-2}$   &  $1.93\times 10^{-3}$	& $5.30\times 10^{-3}$  \\
		$^{197}$Au  &  $3.69\times 10^{-2}$   &  $5.49\times 10^{-3}$	& $1.66\times 10^{-2}$  \\
		$^{208}$Pb  &  $4.38\times 10^{-2}$   &  $6.53\times 10^{-3}$	& $2.21\times 10^{-2}$  \\
		\hline\hline
	\end{tabular}
\end{table}

\begin{table}[!htbp]
	\centering
	\renewcommand{\arraystretch}{1.5}
	\caption{Same as Table~\ref{gluon3xnorm1}, but for nNNPDF3.0 gluon nPDFs.}\label{gluon3xnorm130}
	\begin{tabular}{c| c c}
		\hline\hline
		~~~Nucleus~~~ & ~~$N(\hat p^A_{g,\textrm{poly}})$~~ & ~~$N(\hat p^A_{g,\textrm{can}})$~~ \\
		\hline
		$^{4}$He    &$1.69\times 10^{-3}$  &$1.85\times 10^{-3}$\\
		$^{9}$Be    &$5.97\times 10^{-3}$  &$6.12\times 10^{-3}$ \\
		$^{12}$C    &$8.25\times 10^{-3}$  &8.78$\times 10^{-3}$ \\
		$^{56}$Fe   &2.72$\times 10^{-2}$  &$3.52\times 10^{-2}$ \\
		$^{197}$Au  &4.11$\times 10^{-2}$  &$5.26\times 10^{-2}$ \\
		$^{208}$Pb  &$4.43\times 10^{-2}$  &$6.59\times 10^{-2}$ \\
		\hline\hline
	\end{tabular}
\end{table}
    
\end{widetext}

The results of this work are intriguing which worth further investigation to uncover the profound physical connections between PDFs and nPDFs within the framework of quantum information theory. The behavior of nonperturbative QCD is very complex. In contrast to conventional approaches, the idea of entropy offers a promising avenue for elucidating the fundamental symmetries and underlying dynamics of QCD that are otherwise hidden. 

In addition, it is worth noting that the present analysis is restricted to the EMC region, where the nuclear modification exhibits smooth and monotonic behavior that is well suited to the minimum relative entropy hypothesis employed here. Extending the methodology to the full kinematic range would require addressing the qualitatively different behaviors in the shadowing, antishadowing, and Fermi-motion regions. Such an extension necessitates additional physical inputs---for instance, multiple sets of boundary conditions and constraints from parton momentum sum rules---and is left for future investigation.

%%%%%%%%%%%%%%%%%%%
\section{Concluding remarks}
\label{concludingremarks}
%%%%%%%%%%%%%%%%%%%
In this work, beyond conventional observables, we have introduced the KL divergence, which is a well-known quantity in quantum information theory, as a measure to probe the parton structure of nucleons bound in nuclei. In the quark sector, the resulting structure functions in the EMC region are broadly consistent with current global analyses. In the gluon sector, the minimum relative entropy hypothesis provides an information-theoretic perspective for comparing different global analyses. The present results indicate that, at the level of central values, the EPPS21 gluon distributions are closer to the minimum relative entropy solutions obtained here, while a clearer assessment of nNNPDF3.0 will require further study. These observations suggest that the KL-divergence approach may serve as a useful complementary tool in future studies of nPDFs.

In particle physics, determining nonperturbative inputs beyond global fitting is not only complementary but also necessary for a deeper understanding of hadron structure. The introduction of KL divergence and the minimum relative entropy hypothesis offers a new perspective on longstanding problems. The present framework may help explore further details of nuclear structure functions and parton distributions. Admittedly, the results presented here should still be regarded as indicative rather than definitive, and we hope that this work will stimulate further studies on information-theoretic approaches to nucleon structure.

\section*{Acknowledgements}
The authors would like to thank Prof. Wei Wang and Shuai Zhao for fruitful and inspiring discussions. This work is supported in part by the National Natural Science Foundation of China under Grant No. 12335003, and by the Fundamental Research Funds for the Central Universities under No. lzujbky-2023-stlt01, lzujbky-2024-oy02 and lzujbky-2025-eyt01, the Scientific Research Innovation Capability Support Project for Young Faculty under Grant No. ZYGXQNJSKYCXNLZCXM-P2. J.X is supported in part by the National Natural Science Foundation of China under Grant No. 12475098 and 12105247.

\begin{widetext}
	\appendix
	%%%%%%%%%%%%%%%%%%%%%%%
	\section{Details of the KL-divergence calculations}\label{AppA}
	%%%%%%%%%%%%%%%%%%%%%%%

    Here, we compute the KL divergence using  Eqs.\,(\ref{DKLnor}) and (\ref{DKLfull}). The corresponding integration limits are $[10^{-6}, 0.99]$ and $[0.25, 0.65]$. The resulting KL divergences are designated as $D_{\mathrm{KL}}^{\textrm{full}}(p^A \| q^d)$ and $D_{\mathrm{KL}}^{\textrm{EMC}}(p^A \| q^d)$. Table \ref{KLfullandinter} presents the calculated KL divergences for different nuclei.

\begin{table}[!htbp]
	\centering
	\renewcommand{\arraystretch}{1.5}
	\caption{ The KL divergences for different nuclei in full-$x$ region $[10^{-6},0.99]$ and intermediate-$x$ region $[0.25,0.65]$. The four-momentum transfer is chosen as $Q^2= 10\,\textrm{GeV}^2$.} \label{KLfullandinter}
	\begin{tabular}{c|c c}
		\hline\hline
		~~~Nucleus~~~ & ~ $D_{\mathrm{KL}}^{\textrm{full}}(p^A \| q^d)$ ~ & ~ $D_{\mathrm{KL}}^{\textrm{EMC}}(p^A \| q^d)$ ~ \\
		\hline
		$^{3}$He   &  $1.33 \times 10^{-3}$   &  $0.46 \times 10^{-4}$   \\
		$^{4}$He   &  $5.00 \times 10^{-3}$   &  $2.32  \times 10^{-4}$   \\
		$^{9}$Be   &  $5.22 \times 10^{-3}$   &  $4.75  \times 10^{-4}$   \\
		$^{12}$C   &  $5.42 \times 10^{-3}$   &  $3.70  \times 10^{-4}$   \\
		$^{56}$Fe  &  $6.38 \times 10^{-3}$   &  $8.17  \times 10^{-4}$   \\
		$^{197}$Au &  $7.93 \times 10^{-3}$   &  $16.19 \times 10^{-4}$   \\
		$^{208}$Pb &  $8.03 \times 10^{-3}$   &  $16.85 \times 10^{-4}$   \\
		\hline\hline
	\end{tabular}
\end{table}	
%%%%%%%%%%%%%%%

We use the EPPS21 nPDF set together with the CT18A free-PDF set. For the full-$x$ region $[10^{-6},0.99]$ and the intermediate-$x$ region $[0.25,0.65]$, we take the $F_2^A(x,Q^2)/A$ as $p^A (x)$ in Eq.\,(\ref{SFforA2}) and the $F_2^d(x,Q^2)/2$ as $q^d (x)$. Qualitatively speaking, as $A$ increases, nuclear modification becomes more apparent, leading to an increase in the KL divergence---a trend that is consistent with the results shown in Table \ref{KLfullandinter}. However, it is worth noting that the value for ${^{12}\textrm{C}}$ is lower than for 
	${^{9}\textrm{Be}}$, a pattern consistent with findings in studies of nucleon short-range correlations which show a non-monotonic dependence on 
	A for ${^{12}\textrm{C}}$ and ${^{9}\textrm{Be}}$ \cite{CLAS:2019vsb}. This may arise from the unique nature of beryllium, whose density is similar to helium, yet whose EMC effect closely resembles that of denser nuclei due to its significant cluster structure.
    
The values of normalization factors $z_p$ and $z_q$ defined in Eq.\,(\ref{DKLnor}) are listed in Table \ref{normalization factors}.

	\begin{table}[!htbp]
		\centering
		\renewcommand{\arraystretch}{1.5}
		\caption{The normalization factors $z_p$ and $z_q$ for different nuclei in full-$x$ region $[10^{-6},0.99]$ and intermediate-$x$ region $[0.25,0.65]$. The four-momentum transfer is chosen as $Q^2= 10\,\textrm{GeV}^2$.}\label{normalization factors}
		\begin{tabular}{c|c c}
			\hline\hline
			~~~Nucleus~~~ & ~ $z_p^{\textrm{full}}$ ~ & ~ $z_p^{\textrm{EMC}}$ ~ \\
			\hline
			$^{2}$D($z_q$)    &  $7.06$   &  $19.61 $   \\
			$^{3}$He   &  $6.70$   &  $18.16 $   \\
			$^{4}$He   &  $7.12$   &  $20.32 $   \\
			$^{9}$Be   &  $7.26$   &  $21.05 $   \\
			$^{12}$C   &  $7.14$   &  $20.48 $   \\
			$^{56}$Fe  &  $7.27$   &  $21.16 $   \\
			$^{197}$Au &  $7.48$   &  $22.20 $   \\
			$^{208}$Pb &  $7.50$   &  $22.29 $   \\
			\hline\hline
		\end{tabular}
	\end{table}	
	%%%%%%%%%%%%%%%
	
	Table \ref{gluon normalization factors} presents the normalization factors used in the calculation of the gluon KL divergence.
	
	\begin{table}[!htbp]
		\centering
		\renewcommand{\arraystretch}{1.5}
		\caption{Normalization factors $z^{\textrm{EMC}}_{p,g}$ and $z^{\textrm{EMC}}_{q,g}$ used in the EPPS21 gluon KL-divergence calculation in the intermediate-$x$ region $[0.25,0.5]$.  
        } \label{gluon normalization factors}
		\begin{tabular}{c| c}
			\hline\hline
			~~~Nucleus~~~ & ~ $z_{p,g}^{\textrm{EMC}}$ ~ \\
			\hline
			$^{2}$D($z_{q,g}$)   & $7.67 $   \\
			$^{3}$He   &    $7.74 $   \\
			$^{4}$He   &    $7.93 $   \\
			$^{9}$Be   &    $7.97 $   \\
			$^{12}$C   &    $7.98 $   \\
			$^{56}$Fe  &    $8.09 $   \\
			$^{197}$Au &    $8.20 $   \\
			$^{208}$Pb &    $8.21 $   \\
			\hline\hline
		\end{tabular}
	\end{table}	
	
	Table \ref{nNNPDF-gluon normalization factors} presents the normalization factors used in the calculation of the nNNPDF3.0 gluon KL divergence.
	
	\begin{table}[!htbp]
		\centering
		\renewcommand{\arraystretch}{1.5}
		\caption{Same as Table~\ref{gluon normalization factors}, but for nNNPDF3.0 gluon nPDFs in intermediate-$x$ region $[0.3,0.55]$.} \label{nNNPDF-gluon normalization factors}
		\begin{tabular}{c| c}
			\hline\hline
			~~~Nucleus~~~ & ~ $z_{p,g}^{\textrm{EMC}}$ ~ \\
			\hline
			$^{2}$D($z_{q,g}$)   & $13.03 $   \\
			$^{4}$He   &    $12.99 $   \\
			$^{9}$Be   &    $12.79 $   \\
			$^{12}$C   &    $12.65 $   \\
			$^{56}$Fe  &    $11.94 $   \\
			$^{197}$Au &    $12.32 $   \\
			$^{208}$Pb &    $12.48 $   \\
			\hline\hline
		\end{tabular}
	\end{table}

\section{Collection of determined parameters}\label{AppB}
%%%%%%%%%%%%%%%%%%%%%%%
In this appendix, we collect the determined parameters in the polynomial and canonical parameterizations. For the quark case, we give the four parameters ($a_1$, $b_1$, $c_1$ and $d_1$) in Eq.\,(\ref{polypara}) and four parameters ($a_2$, $b_2$, $c_2$ and $d_2$) in Eq.\,(\ref{canpara}), which are determined by endpoint constraints and minimizing the KL divergences in Eq.\,(\ref{DKLEMC}) for different nuclei.

\begin{table}[!htbp]
	\centering
	\renewcommand{\arraystretch}{1.5}
	\caption{For the quark case, the determined parameters in polynomial form in Eq.\,(\ref{polypara}) and canonical form in Eq.\,(\ref{canpara}) for different nuclei. The integration interval is $x \in [0.25, 0.65]$.}\label{paraab}
	\begin{tabular}{c|c c c c | c c c c}
		\hline\hline
		~~~Nucleus~~~ & ~~~~$a_1$~~~~ & ~~~~$b_1$~~~~ & ~~~~$c_1$~~~~ &~~~~$d_1$~~~~ & ~~~~$a_2$~~~~ & ~~~~$b_2$~~~~ & ~~~~$c_2$~~~~ &~~~~$d_2$~~~~ \\
		\hline
		$^{3}$He    &  $4.29$   &  $-5.39$	& $2.19$  & $0.79$  & $2.99$ & $0.45$  & $-0.90$  & $-2.68$ \\
		$^{4}$He    &  $-5.89$   &  $7.39$	& $-3.00$  & $1.38$  & $0.20$ & $-0.69$  & $1.39$  & $4.13$ \\
		$^{9}$Be    &  $-8.75$   &  $10.97$	& $-4.45$  & $1.54$  & $0.08$ & $-1.06$  & $2.15$  & $6.37$ \\
		$^{12}$C    &  $-7.60$   &  $9.53$	& $-3.86$  & $1.48$  & $0.13$ & $-0.90$  & $1.81$  & $5.37$ \\
		$^{56}$Fe   &  $-11.51$   &  $14.43$	& $-5.85$  & $1.72$  & $0.039$ & $-1.40$  & $2.84$  & $8.43$ \\
		$^{197}$Au  &  $-15.94$   &  $19.97$	& $-8.08$  & $1.98$  & $0.009$ & $-2.04$  & $4.16$  & $12.30$ \\
		$^{208}$Pb  &  $-16.23$   &  $20.32$	& $-8.22$  & $1.99$  & $0.008$ & $-2.09$  & $4.25$  & $12.57$ \\
		\hline\hline
	\end{tabular}
\end{table}

For the gluon case, after applying the EPPS21 global fitting for gluon nPDFs and the minimum relative entropy hypothesis, we give the four parameters ($a_1^g$, $b_1^g$, $c_1^g$ and $d_1^g$) in Eq.\,(\ref{gluonresultspolypara}) and four parameters ($a_2^g$, $b_2^g$, $c_2^g$ and $d_2^g$) in Eq.\,(\ref{gluonresultscanpara}), which are determined by endpoints constraints and minimizing the KL divergences in Eq.\,(\ref{DKLEMCgluon}) for different nuclei.

\begin{table}[!htbp]
	\centering
	\renewcommand{\arraystretch}{1.5}
	\caption{For the gluon case, the determined parameters in polynomial form in Eq.\,(\ref{gluonpolypara}) and canonical form in Eq.\,(\ref{gluoncanpara}) for different nuclei by utilizing the EPPS21 global fits. The integration interval is $x \in [0.25, 0.50]$.}\label{gluonparaab}
	\begin{tabular}{c|c c c c | c c c c}
		\hline\hline
		~~~Nucleus~~~ & ~~~~$a_1^g$~~~~ & ~~~~$b_1^g$~~~~ & ~~~~$c_1^g$~~~~ &~~~~$d_1^g$~~~~ & ~~~~$a_2^g$~~~~ & ~~~~$b_2^g$~~~~ & ~~~~$c_2^g$~~~~ &~~~~$d_2^g$~~~~ \\
		\hline
		$^{3}$He    &  $-8.09$   &  $8.57$	& $-2.98$  & $1.33$     & $0.20$  & $-0.65$  & $2.29$  &  $5.43$  \\
		$^{4}$He    &  $-29.62$   &  $31.38$	& $-10.89$  & $2.22$   & $0.002$  & $-2.56$  & $8.90$  &  $21.19$ \\
		$^{9}$Be    &  $-34.85$   &  $36.92$	& $-12.81$  & $2.44$   & $0.002$  & $-2.49$  & $9.06$  &  $21.22$ \\
		$^{12}$C    &  $-36.95$   &  $39.15$	& $-13.58$  & $2.52$   & $0.0004$  & $-3.22$  & $11.20$  &  $26.68$ \\
		$^{56}$Fe   &  $-50.10$   &  $53.05$	& $-18.40$  & $3.06$   & $0.003$  & $-2.37$  & $9.83$  &  $21.94$ \\
		$^{197}$Au  &  $-60.90$   &  $67.63$	& $-23.44$  & $3.63$   & $0.006$  & $-2.00$  & $9.40$  &  $20.05$ \\
		$^{208}$Pb  &  $-64.57$   &  $68.34$	& $-23.69$  & $3.66$   & $0.002$  & $-2.36$  & $10.92$  &  $23.50$ \\
		\hline\hline
	\end{tabular}
\end{table}

For the gluon case, after applying the nNNPDF3.0 global fitting for gluon nPDFs and the minimum relative entropy hypothesis, we give the four parameters ($a_1^g$, $b_1^g$, $c_1^g$ and $d_1^g$) in Eq.\,(\ref{gluonresultspolypara3.0}) and four parameters ($a_2^g$, $b_2^g$, $c_2^g$ and $d_2^g$) in Eq.\,(\ref{gluonresultscanpara3.0}), which are determined by endpoints constraints and minimizing the KL divergences in Eq.\,(\ref{DKLEMCgluon}) for different nuclei.

\begin{table}[!htbp]
	\centering
	\renewcommand{\arraystretch}{1.5}
	\caption{Same as Table~ \ref{gluonparaab}, but for nNNPDF3.0 gluon nPDFs in intermediate-$x$ region $[0.3, 0.55]$. }\label{gluonparaab3.0}
	\begin{tabular}{c|c c c c | c c c c}
		\hline\hline
		~~~Nucleus~~~ & ~~~~$a_1^g$~~~~ & ~~~~$b_1^g$~~~~ & ~~~~$c_1^g$~~~~ &~~~~$d_1^g$~~~~ & ~~~~$a_2^g$~~~~ & ~~~~$b_2^g$~~~~ & ~~~~$c_2^g$~~~~ &~~~~$d_2^g$~~~~ \\
		\hline
		$^{4}$He    &  $-8.97$   &  $10.96$	& $-4.41$  & $1.60$   & $0.0976$  & $-1.03$  & $2.22$  &  $6.33$ \\
		$^{9}$Be    &  $-32.70$   &  $39.96$	& $-16.06$  & $3.17$   & $0.0060$  & $-2.24$  & $5.44$  &  $14.74$ \\
		$^{12}$C    &  $-47.08$   &  $57.52$	& $-23.11$  & $4.13$   & $0.0014$  & $-2.86$  & $7.16$  &  $19.15$ \\
		$^{56}$Fe   &  $-216.67$   &  $263.53$	& $-105.39$  & $15.07$   & $0.0025$  & $-2.20$  & $15.47$  &  $30.52$ \\
		$^{197}$Au  &  $-359.78$   &  $435.71$	& $-173.42$  & $23.93$   & $1.58*10^{-6}$  & $-4.96$  & $30.91$  &  $62.49$ \\
		$^{208}$Pb  &  $-389.23$   &  $470.88$	& $-187.19$  & $25.70$   & $0.0002$  & $-2.60$  & $31.26$  &  $56.80$ \\
		\hline\hline
	\end{tabular}
\end{table}

\end{widetext}

\end{document}